\documentclass[%
reprint,
superscriptaddress,
longbibliography,
amsmath,amssymb,
aps,
prb,
floatfix,
]{revtex4-2}

\usepackage{graphicx}%
\usepackage{xcolor}
\usepackage{dcolumn}%
\usepackage{bm}%
\usepackage[
colorlinks,
linkcolor=blue,
citecolor=blue,
urlcolor=blue
]{hyperref}%
\usepackage[
subrefformat=parens,
labelformat=parens,
caption=false
]{subfig}
\usepackage{booktabs}

\DeclareMathOperator{\sgn}{sgn}

\begin{document}
\title{
Quantum Hall interferometry at finite bias with multiple edge channels 
}
\author{Zezhu Wei}
\author{D. E. Feldman}
\affiliation{Department of Physics, Brown University, Providence, Rhode Island 02912, USA}
\affiliation{Brown Theoretical Physics Center, Brown University, Providence, Rhode Island 02912, USA}
\author{Bertrand I. Halperin}
\affiliation{Department of Physics, Harvard University, Cambridge, Massachusetts 02138, USA}

\date{\today}

\begin{abstract}
In a quantum Hall interferometer, the dependence of the signal on source-drain voltage is controlled by details of the edge physics, such as the velocities of edge modes and the interaction between them and with screening layers. Such dependence of the signal has been seen in recent experiments at various integer and fractional filling factors, including $\nu=2$ and $\nu=2/5$, where two edge modes are present. Here we study theoretically the current-voltage curves for various values of the relative edge velocities, interaction strength, and the temperature, in a model containing two edge modes. We consider separate cases in which the inner mode or the outer mode is weakly backscattered at the tunneling contacts. When the inner mode is completely reflected and the outer mode is partially transmitted, we find striking features at very low temperatures related to resonance of excitations of the closed inner channel. Fluctuations in the charge of the closed inner mode, caused by sparse tunneling events, lead to an exponential suppression of the interference visibility at high voltages, in agreement with experiments.
\end{abstract}

\maketitle

\section{Introduction}

Interferometric methods have proven to be a powerful experimental tool for the investigation of quantum Hall systems \cite{MZ-heiblum,camino2007:PhysRevLett.98.076805,willett09:pnas.0812599106,zhang2009:PhysRevB.79.241304,mcclure2009:PhysRevLett.103.206806,ofek2010role,mcclure2012:PhysRevLett.108.256804,manfra19,manfra20,deprez2021tunable,ronen2021,zhao2022graphene,nakamura2023:fabry,Kim,chamon1997:PhysRevB.55.2331,dassarma2005:PhysRevLett.94.166802,stern2006:PhysRevLett.96.016802,bonderson2006:eo2,MZ2,law2006:PhysRevB.74.045319,halperin2011:PhysRevB.83.155440,smits2014:PhysRevB.89.045308,ferraro2017:10.21468/SciPostPhys.3.2.014,review-FH,fh2022}. Most dramatically, for fractional quantized Hall states, interferometry has provided direct measurements of the fractional statistics as well as the fractional charges of quasiparticle excitations.   More generally, for both integer and fractional quantum Hall states, interferometry has provided information about the nature of the modes that propagate along the  edges of the interferometer, including their interactions with localized charges in the bulk.  At a higher level, interferometer experiments  provide some sensitive tests of our understanding of quantum Hall states in real materials. 

In this paper, we restrict ourselves to quantum Hall interferometers of the Fabry-P\'erot type \cite{chamon1997:PhysRevB.55.2331}. In this geometry, two constrictions  allow charged quasiparticles or electrons to tunnel between the chiral edge modes on the two sides of a quantum Hall region (Fig.~\ref{fig:FP_interferometer}). When a voltage difference is applied between the two edges of the device, any particles that tunnel  across the constrictions will reduce the current transmitted along the edges, and this will lead to a decrease in the measured electrical conductance. If quasiparticles can propagate coherently along the edges between the two constrictions, the tunneling amplitudes will add coherently with a phase difference that depends on the magnetic flux passing through the interferometer region, with possible additional contributions due to quantum statistics when fractionally-charged quasiparticles are involved. As the enclosed flux is sensitive to relatively small changes in the applied magnetic field or to changes in the area of the device caused by changes in the voltage applied to nearby gates,  changes in these parameters can lead to interference oscillations in the electrical resistance of the device.  It should be noted that changes in the interferometer area can result from Coulomb interactions between the edge modes and charges in the bulk, as well as from direct interactions with the gates \cite{halperin2011:PhysRevB.83.155440}. 

Experiments are most frequently carried out in the limit of small  source-drain voltages, where the transmitted current is a linear function of the applied voltage.  Further information can be obtained, however,   from measurements at larger voltages.  Voltage-dependencies in the case in which the quantized Hall state has a single propagating mode along each edge, such as at filling factors $\nu = 1$ or $\nu = 1/3$, were discussed in Ref.~\cite{chamon1997:PhysRevB.55.2331}. The analysis predicted an oscillatory dependence of the interference amplitude on the applied voltage, with a voltage scale determined by the edge-mode velocity and the length of the interference path.  The precise form of the voltage dependence  was found to depend on the filling factor $\nu$.

In this paper, we extend the theory to cases in which there are two co-propagating modes on each edge. Specifically, we consider a fractional case with bulk filling factor $\nu= 2/5$ and the integer case $\nu = 2$. We take into account the Coulomb coupling between charge fluctuations along the two modes on a given edge, but we assume that over longer distances interactions are well screened, due to the presence of a nearby conducting gate. We also assume weak scattering of charges between the two edge  modes.  This assumption is justified in the case of $\nu=2$, because the associated particles have different spins in the two modes, and tunneling requires a spin flip.
For $\nu=2/5$,  the experiments of Ref.~\cite{nakamura2023:fabry} suggest that intermode tunneling is weak in that case as well, perhaps because of a relatively large separation of the two edge channels and a small interferometer size
(c.f., Ref.~\cite{csg}).
At the same time, inter-mode scattering cannot be ignored on the laboratory time scale. We will see that such rare scattering events can dramatically suppress the observed time-averaged interference current in the non-linear regime.

When the interferometer edge contains two co-propagating modes, different interference phenomena can occur depending on the nature of the constrictions. If the constrictions are only weakly pinched off,   tunneling will occur only between the inner  modes of the two edges, while the outer modes pass freely through the constrictions [see Fig.~\ref{fig:FP_interferometer}(a)]. We shall see that in this case, the presence of the outer mode has relatively little effect on the interference signal, and the voltage-dependence of the interference amplitude is similar to what one would predict for the inner mode alone, using the formulas of Ref.~\cite{chamon1997:PhysRevB.55.2331}. The situation is more interesting  in the regime where the constrictions are pinched to the extent that the inner mode is completely reflected, while the outer mode passes through the constrictions with only weak backscattering [see Fig.~\ref{fig:FP_interferometer}(b)].
This physical situation was previously addressed for $\nu=2$ and $T=0$ under the assumption of an infinite edge velocity \cite{ferraro2017:10.21468/SciPostPhys.3.2.014}. We go well beyond that limiting case.

In another departure  from the earlier work, we consider asymmetric voltage bias. It is usually assumed that the voltage bias is applied symmetrically, that is, the potentials $V_d$ and $V_u$ of the lower and upper edges equal $\pm V/2$, where $V$ is the bias. In actual experiments \cite{Kim} the bias is often asymmetric. We thus introduce the  bias voltage $V=V_d-V_u$ and the average voltage $\tilde V=(V_u+V_d)/2$, which can often be treated as independent parameters. The degree of asymmetry is measured by the parameter $\eta=\tilde V/V$, which ranges from $0$ to $1/2$. Note that for a linear relation $\tilde V=\eta V$ between the bias and average voltages, the conductance depends on the derivatives of the current with respect to both $V$ and $\tilde V$.

Experiments in the case of $\nu=2/5$ appear to be in a regime where the Coulomb coupling between the two edge modes is relatively weak \cite{nakamura2023:fabry}.
On the other hand, in the integer case, experiments can fall in a regime where the two modes are strongly coupled.  Such strong coupling has been proposed \cite{tpair} to explain the  ``pairing'' phenomenon \cite{pair1,pair2} in GaAs systems, where the oscillation period in a range of filling factors above $\nu=2$ coincides with what one might expect for particles of charge $2e$,  and it has been observed directly in a graphene interferometer \cite{Kim, sacepe}. Consequently, in our numerical examples, we shall focus on the strong-coupling regime for $\nu=2$ but the weak coupling regime for $\nu=2/5$.

The signal measured in a quantum Hall interferometer will generally  depend on the number of quasiparticles inside the interference loop. In fractional quantum Hall states, there will be phases associated with quantum statistics of the quasiparticles, in addition to effects of electrostatic interactions, which may be present in integer as well as fractional cases.
For bulk quasiparticles, the latter effect may be neglected if the bulk-edge interaction is well screened, as we assume
in this paper. 
However, electrostatic coupling cannot be neglected between the two copropagating edge modes,  which are close together in space. In the case of outer-mode interference, the inner mode forms a closed loop, which contains an integer number of quasiparticles, and we have to account for possible fluctuations of this number.  

In the linear transport regime, all fluctuations are thermal and become unimportant in the low-temperature situation that we consider in this paper. In the nonlinear transport regime, however, it is possible for charge to tunnel between the inner loop and the outer edge channels, so that a small leakage current flows through the device. Although  the tunneling time scale is much longer that the travel time of a charge through the interferometer along its edges, the leakage current can cause fluctuations in the charge on the closed inner loop on the time scale of an experimental measurement.  As the charge fluctuations grow with increasing voltage bias, the resulting fluctuations in the interference phase can lead to 
rapid suppression of the observed signal at large voltages.

In a separate effect, even when the charge on the closed inner loop is fixed, there can be resonances in the transport when the bias voltage times the tunneling charge matches the energy of a plasma excitation on the inner loop.
 
The organization of our paper is as follows. In Section~\ref{sec:model},  we introduce the models used in our paper, and we outline our computational methods.
In Section~\ref{sec:filling_two}, we  consider in detail the integer quantum Hall effect at $\nu=2$, while in Section~\ref{sec:filling_two_fifth}, we  consider $\nu=2/5$.
In all cases, when we consider interference of the outer mode, we begin with a discussion  under the assumption of a  fixed charge on the island formed by the closed inner channel. In that regime, tunneling between co-propagating edge channels can be neglected. We then address the regime of the fluctuating island charge. This amounts to averaging the results for the fixed charge over the statistical distribution of the charge on the inner edge channel. The fixed charge approximation applies in two cases: when the time-scale of the experiment is shorter than the time over which the island charge changes, and if the island charge does not fluctuate at all, as may be the case in the linear transport regime.

In Secs.~\ref{sec:model} and \ref{sec:filling_two}, we separately consider the cases of asymmetric and symmetric voltage bias.   Since experiments \cite{nakamura2023:fabry} at $\nu=2/5$ were performed at an approximately symmetric voltage bias, however, we assume that the bias is symmetric in the discussion of $\nu=2/5$.

Our findings are summarized in Section~\ref{sec:conclusion}.
Some technical details are addressed in Appendices~\ref{appendix:series} and \ref{appendix:resonance}. 
Appendix~\ref{appendix:additional_phase}  contains a careful calculation of the Aharonov-Bohm phase. Experimental data are usually presented
as a Fourier transform of the interference contribution to the current. We address the subtleties of such a Fourier transform in Appendices  \ref{appendix:glitch} and \ref{appendix:Fourier}.
Appendix \ref{appendix:notations} contains a table of notations in this paper.

\section{Model}\label{sec:model}

\subsection{Edge modes}

We start with the model of a quantum Hall (QH) liquid with two co-propagating chiral edge modes. The outer mode separates the filling factor $\nu_1$ from 0, and the inner mode separates the filling factor $\nu_1$ from the filling factor 
$ \nu = \nu_2+\nu_1$. One example from the integer QH effect (IQHE) is the $\nu=2$ liquid, where $\nu_1=\nu_2=1$; another example from the fractional QH effect (FQHE) is the $\nu=2/5$ liquid, where $\nu_1=1/3$ and $\nu_2=1/15$. The Lagrangian of a single chiral left-moving QH edge in the absence of the intermode interaction is
\begin{align}
\label{1}
	L = {} & \frac{ \hbar \partial_{x} \phi_{\rm o} (\partial_{t}-v_{\rm o} \partial_{x}) \phi_{\rm o}}{4\pi}  + \frac{ \hbar \partial_{x} \phi_{\rm i} (\partial_{t}-v_{\rm i} \partial_{x}) \phi_{\rm i}}{4\pi}  ,
\end{align}
where $v_{\rm o}$ and $v_{\rm i}$ are the velocities of the outer and inner modes.
To describe the opposite propagation direction, one needs to change the sign in front of the time derivatives. The Bose fields $\phi$ describe the linear charge densities on the edge channels, $-\frac{e\sqrt{\nu_i}\partial_x\phi}{2\pi}$,  for a left-moving channel with filling factor discontinuity $\nu_{i}$.
The charge density of a right-moving channel is defined with  the opposite sign: $\frac{e\sqrt{\nu_i}\partial_x\phi}{2\pi}$. Here, $-e<0$ is the electron charge and we have assumed that the magnetic field points in the negative $z$-direction. 
\begin{figure}[!htb]
	\centering
	\subfloat[\label{subfig:FP_interferometer:a}]{%
		\includegraphics[width=\columnwidth]{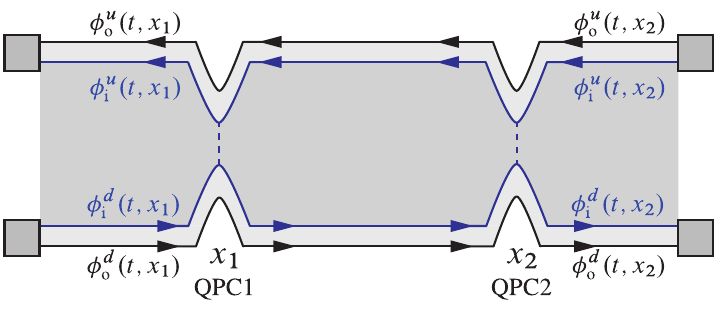}%
	}
	
	\subfloat[\label{subfig:FP_interferometer:b}]{%
		\includegraphics[width=\columnwidth]{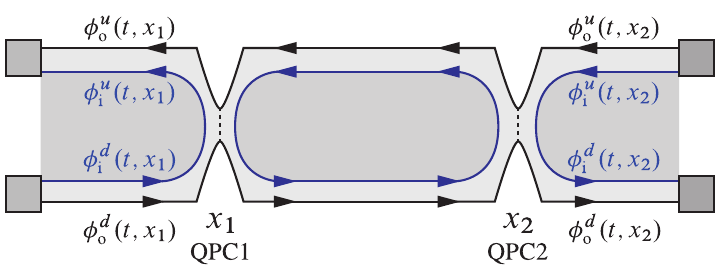}%
	}
	\caption{Schematics of an electronic Fabry-P\'erot interferometer with filling factor $\nu=2/5$, showing inner mode (a) and outer mode (b) tunneling. Tunneling is shown with dashed lines. On each edge of the interferometer, there are two co-propagating edge modes.
 }
	\label{fig:FP_interferometer}
\end{figure}

In contrast to the case of a single edge mode, a Fabry-P\'erot interferometer can now be operated in two ways: inner mode tunneling in Fig.~\ref{fig:FP_interferometer}(a) and outer mode tunneling in Fig.~\ref{fig:FP_interferometer}(b).
In both cases, we assume that the distances between two quantum point-contacts (QPC1 and QPC2) are the same along the two edges, and the coordinates of QPC1 and QPC2 are $x_{1}$ and $x_{2}$. 
The distance is denoted as $a=|x_{2}-x_{1}|$. An adjustment for an asymmetric device with the different distances is straightforward. Fig.~\ref{fig:FP_interferometer}(a) shows the situation of inner mode tunneling, where tunneling occurs between the counter-propagating  modes on the upper ($u$) and lower ($d$) edges at the two QPCs.
Fig.~\ref{fig:FP_interferometer}(b) shows the situation of outer mode tunneling, where the inner bulk region is completely pinched off by the side gates, and the inner mode forms a closed loop within the interference loop, leading to the direct tunneling through the outer bulk.
Inner mode tunneling has been extensively studied in the past (see, in particular, Ref. \cite{smits2014:PhysRevB.89.045308}), and we address that case only briefly. The main focus of the paper is the technically harder and physically richer case of outer mode tunneling.

The Lagrangian (\ref{1}) is written in a gauge such that the vector potential of the magnetic field has zero component  parallel to the edges away from the tunneling contacts. At the contacts, the vector potential is large and points across the constriction. This leads to an Aharonov-Bohm phase in the tunneling operator, which depends on the magnetic field, the charge of the tunneling particle, and the area enclosed by the interferometer path. For fractional Hall states, there can be  additional contributions to the phase, arising from the braiding statistics with quasiparticles inside the loop.  In the case of a closed inner channel,  the total phase accumulation around the loop, including the Aharonov-Bohm phase and any braiding phases due to enclosed anyons, is required to obey a quantization condition, which  leads to quantization of the total charge on the closed loop (see subsection \ref{jumps}).

We note that the fields $\phi$ in Eq. (\ref{1}) and the associated charge densities should be understood to describe fluctuations or deviations from an equilibrium state having a specified value of the voltages applied to any nearby gates, as well as fixed voltage on the edge states and a specified magnetic field.  Varying voltages from these fiduciary values will lead to additional terms in the Lagrangian, linear in the gradients $\partial_x \phi$.

Obviously, the noninteracting model (\ref{1}) is oversimplified. 
A more realistic model adds a short-range Coulomb interaction to the Lagrangian~\cite{ferraro2017:10.21468/SciPostPhys.3.2.014}, 
\begin{equation}
	L_e  = - \frac{\hbar w}{2\pi} (\partial_{x} \phi_{\rm o})(\partial_{x}\phi_{\rm i}),
\end{equation}
where the Coulomb repulsion strength $w>0$.
The total Lagrangian $L+L_{e}$ can be diagonalized by introducing the following transformation,
\begin{equation}\label{eq:eigen_modes}
	\begin{pmatrix}
		\phi_{\rm o}	\\
		\phi_{\rm i}	
	\end{pmatrix}	= 
	\begin{pmatrix}
		\cos\theta	& -\sin\theta \\
		\sin\theta & \cos\theta
	\end{pmatrix}
	\begin{pmatrix}
		\phi_{1}	\\
		\phi_{2}	
	\end{pmatrix}
\end{equation}
where $\phi_{1,2}$ are the edge eigenmodes. Having in mind $\nu=2/5$, we expect that the bare velocity $v_{\rm o}$ of the outer $1/3$ mode is faster than the velocity $v_{\rm i}$ of the inner $1/15$ mode.
If we choose $0<\theta<\pi/4$ satisfying $\tan(2\theta) = 2 {w} /(v_{\rm o}-v_{\rm i}) $, then the diagonalized Lagrangian is
\begin{equation}
	L+L_{e} = \frac{ \hbar\partial_{x} \phi_{1} (\partial_{t}-v_{1} \partial_{x}) \phi_{1}}{4\pi} +
	\frac{\hbar \partial_{x} \phi_{2} (\partial_{t}-v_{2} \partial_{x}) \phi_{2}}{4\pi}  ,
\end{equation}
with the velocities of the eigenmodes being
\begin{align}\label{eq:eigen_velocity}
	v_{1} &= \frac{1}{2} \left[ v_{\rm o} + v_{\rm i} + \sqrt{(v_{\rm o}-v_{\rm i})^2+4 w^2}  \right],\\
	v_{2} &= \frac{1}{2} \left[v_{\rm o} + v_{\rm i} - \sqrt{(v_{\rm o}-v_{\rm i})^2+4 w^2}\right]  .
\end{align}

In the following sections, we shall deal with the inner and outer mode tunneling in the presence of short-range interaction. 
We discuss tunneling contacts in subsection \ref{sec:II_B}. Subsection \ref{sec:II_C} deals with the case of the open inner channel. Subsections \ref{sec:II_D} and \ref{sec:II_E} address a much harder problem of the closed inner channel.
We make the assumption that the interaction strength $w$ is less than $\sqrt{v_{\rm i} v_{\rm o}}$ and hence $v_{2}>0$
as thermodynamic stability demands.

As mentioned above, we assume weak tunneling between the modes on a single edge. %
Nevertheless, rare tunneling events are important at a high bias. We thus approach the problem in two steps. First, we neglect inter-mode tunneling in subsection \ref{sec:II_D}. This amounts to computing the tunneling current through the interferometer on a short time scale. Next, we introduce inter-mode tunneling between co-propagating channels in subsection \ref{sec:II_E}. We only consider inter-mode tunneling events in the closed-loop geometry \ref{fig:FP_interferometer}(b) since only in that case is there long-term memory of rare tunneling events due to the charge conservation on the closed loop. The memory effect means that the  interference current, computed at a fixed charge of the inner island, should be averaged over all possible chargers on the inner loop. 

In principle, tunneling between the edges and localized states in the bulk of the interferometer can affect the current even in geometry \ref{fig:FP_interferometer}(a), where both channels are open. We will not consider this part of physics in this paper. Such tunneling was addressed in the linear transport regime at finite temperatures in Ref.~\cite{halperin2011:PhysRevB.83.155440}. Our analysis should apply, however, if the sample is in an incompressible state, with at most a few localized states inside the energy gap in the interior of the interferometer. 

Most of our analysis assumes symmetric voltage bias, $\tilde V=0$. Yet, the experiment \cite{Kim} at $\nu=2$ was performed at asymmetric bias. We extend the theory to the asymmetric case in subsection \ref{sec:filling_two_asymmetric}.

\subsection{Quantum point contacts}
\label{sec:II_B}

In general, for a two-point Fabry-P\'erot interferometer where quasiparticles with charge $e^{*}$ tunnel and a voltage bias $V$ is applied, the total Hamiltonian is $\hat{H}+\hat{H}_{T}$, where $\hat{H}$ describes the edges and $\hat{H}_T$ describes the interedge tunneling. The edge Hamiltonian $\hat{H}$ is the sum of the Hamiltonians of the upper and lower edges,
\begin{align}
    \hat{H} = \int dx \,\left\{ \frac{\hbar v_{1}}{4\pi} [(\partial_{x} \phi_{1}^{u})^2 + (\partial_{x} \phi_{1}^{d})^2]
    + \frac{\hbar v_{2}}{4\pi} [(\partial_{x} \phi_{2}^{u})^2 + (\partial_{x} \phi_{2}^{d})^2] \right\} ,
\end{align}
where the indexes $u$ and $d$ refer to the top and bottom edges.
As shown in Fig.~\ref{fig:FP_interferometer}, we choose the lower edge to be right-moving and the upper edge to be left-moving; therefore, their Bose fields satisfy the following commutation relation~\cite{vondelft1998:bosonization},
\begin{align}
	[ \phi^{d/u}_{j}(t,x), \phi^{d/u}_{k}(t,y) ] & = \pm i \pi \sgn(x-y) \delta_{jk},
\end{align}
where the plus sign corresponds to the lower edge ($d$) and the minus sign corresponds to the upper edge ($u$).
The tunneling Hamiltonian $\hat{H}_{T}$ is expressed as
\begin{equation}
	\hat{H}_{T} = \Gamma_{1}  \hat{T}(t,x_{1}) + \Gamma_{2}  \hat{T}(t,x_{2}) + \mathrm{H.c.},
\end{equation}
where $\hat{T}(t, x)$ are the tunneling operators that transfer particles from the lower edge to the upper edge through the QPCs, $x_{1,2}$ are the locations of the QPCs, and $\Gamma_{1,2}$ are the tunneling amplitudes. Voltage bias enters as a difference of the chemical potentials of the edges. It can be eliminated in a usual way~\cite{law2006:PhysRevB.74.045319} in the interaction representation so that the tunneling Hamiltonian becomes
\begin{equation}
\label{HT}
	\hat{H}_{T} = \Gamma_{1} e^{i \omega_{J} t} \hat{T}(t,x_{1}) + \Gamma_{2} e^{i \omega_{J} t} \hat{T}(t,x_{2}) + \mathrm{H.c.},
\end{equation}
where $\omega_{J} = e^{*}V/\hbar$ is the effective Josephson frequency, 
$e^*$ is the charge of the tunneling particle, and $V \equiv V_d - V_u$ is the bias voltage, i.e., the difference between voltages applied to the lower and upper edges. The bias voltage affects the average charge density on the edges. For an asymmetrically applied voltage bias, where $V_u+V_d \neq 0$, this may alter the net charge on the edge and  induce a voltage-dependent shift in the interference phase~\cite{law2006:PhysRevB.74.045319}.  This effect will be discussed in subsection 
\ref{sec:filling_two_asymmetric}  below.

The tunneling current operator is the time derivative of the charge of one edge, and it can be represented as 
\begin{equation}
	\hat{I}_{T} = \frac{i e^*}{\hbar} [\Gamma_{1} e^{i \omega_{J} t} \hat{T}(t,x_{1}) + \Gamma_{2} e^{i \omega_{J} t} \hat{T}(t,x_{2}) - \mathrm{H.c.}].
\end{equation}
Throughout this paper we will work in the weak backscattering regime, i.e., we assume a small tunneling amplitude for each quantum point contact and treat the tunneling Hamiltonian $\hat{H}_{T}$ as a perturbation.
In the lowest-order perturbation theory, the expectation value for the current is found using the following standard formula,
\begin{equation}
	I_{T}  = -\frac{i}{\hbar} \int_{-\infty}^{0} d{t} \langle{[ \hat{I}_{T}(0), \hat{H}_{T}(t) ]}\rangle.
\end{equation}
By defining
\begin{equation}
\label{13}
	P(t,x) = \langle \hat{T}(t,x) \hat{T}^{\dagger}(0,0) \rangle
	= \langle \hat{T}^{\dagger}(t,x) \hat{T}(0,0) \rangle,
\end{equation}
which satisfies the relation $P^{*}(t,x) = P(-t,-x)$, one can write down the non-interference part of the tunneling current $I_T=I_{\text{non-int}}+I_{\text{int}}$ as
\begin{align}\label{eq:non-int_general}
	I_{ \text{non-int}} = \frac{e^{*}}{\hbar^2}	\int_{-\infty}^{0} d{t} (|\Gamma_{1}|^2+|\Gamma_{2}|^2) (e^{-i \omega_{J}t}-e^{i\omega_{J}t}) \nonumber\\
	\times [ P(-t,0)-P(t,0) ],
\end{align}
and the interference part as
\begin{align}\label{eq:int_general}
	I_{ \text{int}} =  \frac{e^{*}}{\hbar^2} \int_{-\infty}^{0} d{t} (  \Gamma_{1} \Gamma_{2}^{*} e^{-i \omega_{J}t} - \mathrm{c.c} )
	[P(-t, -a) - P(t,a) ]\nonumber\\
	+ ( \Gamma_{1}^{*} \Gamma_{2} e^{-i \omega_{J} t} -  \mathrm{c.c}) [ P(-t,a) - P(t,-a) ],
\end{align}
where $a=|x_{1}-x_{2}|$ is the distance between the two QPCs, measured along an edge.

In the next two subsections, we address the structure of the correlation functions for inner and outer mode tunneling. Another component of the model is the phase difference $\varphi$ between the two tunneling amplitudes.
Its behavior is rather subtle for outer mode tunneling. We discuss it in subsection \ref{sec:II_E}.

\subsection{Inner mode tunneling}
\label{sec:II_C}

In this section, we discuss the inner mode tunneling case shown in Fig.~\ref{fig:FP_interferometer}(a), with quasiparticle charge 
$e^{*} = - \nu e/2$, which is the smallest allowed charge for $\nu=2/5$ or $\nu=2$. The quasiparticle charge can also be written as $e^* = - n \nu_2 e$, where $n$ is an integer, since $\nu/2\nu_2$ is  an integer.
We address inner mode tunneling only briefly, as a more detailed investigation of inner mode tunneling can be found in Ref. \cite{smits2014:PhysRevB.89.045308}.

According to the bosonization technique, the tunneling operator
\begin{equation}
    \label{Top}   
    \hat{T}(t,x) 
    = \delta^{-n^2\nu_2} e^{ i n \sqrt{\nu_2} \phi_{\rm i}^{u}(t,x)} e^{- i n \sqrt{\nu_2} \phi_{\rm i}^{d}(t,x)},
\end{equation}
where $\delta$ is a short-time cutoff. We will denote the correlation function (\ref{13}) as $P_{\rm i}$ since we deal with the inner modes.
 
Using standard formulas for correlation functions of Bose fields, it is easy to find that at zero temperature,
\begin{equation}\label{eq:P(t,x)_inner}
	P_{\rm i}(t,x) = G_{d}(t,x) G_{u}(t,x),
\end{equation}
where
\begin{align}
	G_{d,u}(t,x) = {}& [ \delta + i (t \mp x/v_{1}) ]^{-n^2\nu_2  \sin^2\theta } \nonumber\\
    &\times [ \delta + i (t \mp x/v_{2}) ]^{-n^2\nu_2\cos^2\theta}.
\end{align}
We hence obtain an analytical result for the non-interference current as follows:
\begin{equation}
	I_{ \rm i, \text{non-int}} = \frac{2\pi e^{*}( |\Gamma_{1}|^2 + |\Gamma_{2}|^2 ) }{\Gamma(2 n^2 \nu_2) \hbar^2}  |\omega_{J}|^{2n^2\nu_2-1}
	\sgn(\omega_{J}),
\end{equation} 
where $\Gamma(2 n^2 \nu_2)$ in the denominator is the Gamma function.
For the interference current, we first introduce $\exp(i\varphi)$ as the phase difference between $\Gamma_{1}$ and $\Gamma_{2}$, i.e., $\Gamma_{2}/\Gamma_{1} = |\Gamma_{2}/\Gamma_{1}| \exp(i\varphi)$.
The phase
difference includes an Aharonov-Bohm phase, the statistical phase due to any localized anyons, a non-universal phase due to microscopic details of the tunneling contacts, and a possible correction due to asymmetry of the voltage bias 
(see Section \ref{sec:filling_two}). 
Since interference experiments are conducted in a regime where the interferometer contains a large number of flux quanta, a relatively small change in magnetic field or gate voltage can alter the Aharonov-Bohm phase by a large amount, while phase shifts due to changes in the contacts remain negligible. In our discussions, below, we also assume that there are no changes in the number of localized anyons produced by changes in the applied voltages or magnetic fields.

 Noticing that $P_{\rm i}(t,x) = P_{\rm i}(t,-x)$, we can write the expression for $I_{\rm i, \text{int}}$ as,
\begin{equation}
	\frac{2e^{*}}{\hbar^2} |\Gamma_{1} \Gamma_{2}| \cos\varphi \int_{-\infty}^{0} dt
	\left[ e^{-i\omega_{J}t} P(-t,a) - e^{i\omega_{J}t} P(-t,a) + \text{c.c.}  \right].
\end{equation}
At a finite temperature $T$, we need to replace the zero-temperature correlation functions with the finite-temperature ones. This can be done by a conformal transformation~\cite{chamon1995:PhysRevB.51.2363}, and the result is
\begin{align}
    \label{eq:conformal}
	G_{d,u}(t,x) = (\pi T/\hbar)^{n^2 \nu_2} [ \sin\pi T( \delta + i (t \mp x/v_{1}))/\hbar ]^{- n^2 \nu_2 \sin^2\theta} \nonumber\\
	\times[ \sin \pi T (\delta + i (t \mp x/v_{2})) /\hbar ]^{- n^2 \nu_2 \cos^2\theta},
\end{align}
where we have taken $k_{B}=1$.

\subsection{Outer mode tunneling: fixed island charge}
\label{sec:II_D}

In this section, we discuss the geometry in Fig.~\ref{fig:FP_interferometer}(b) for the outer mode tunneling, where the tunneling charge $e^{*} = -m \nu_1 e$, the tunneling operator $\hat{T}(t,x)$ is $\delta^{-m^2 \nu_1} e^{ i m \sqrt{\nu_1} \phi_{\rm o}^{u}(t,x)} e^{-i m \sqrt{\nu_1} \phi_{\rm o}^{d}(t,x)}$, and $\delta$ is the ultra-violet cutoff. In the subsequent sections, we will focus on the lowest-charge quasiparticles with $m=1$, but the equations in this sections are general.
In contrast to the inner mode tunneling, due to the presence of the inner closed loop, the correlation function (\ref{13}), $P_{\rm o}(t,x)$, does not have a simple form as $P_{\rm i}(t,x)$ in Eq.~(\ref{eq:P(t,x)_inner}).
Indeed, even in the absence of charge tunneling between the two edge modes,  the Coulomb interaction  between the closed inner channel and the outer $\nu_1$ channel affects the amplitudes for scattering of the two incoming $\nu_1$ edge modes, $\phi_{\rm o}^{d}(\omega,x_{1})$ and $\phi_{\rm o}^{u}(\omega,x_{2})$, into the outgoing modes, $\phi_{\rm o}^{u}(\omega,x_1)$ and $\phi_{\rm o}^{d}(\omega,x_2)$.
That scattering process is the focus of the current subsection, as 
we neglect inter-mode tunneling between the inner and outer channels, and we consider the total  charge on the inner mode to be fixed.

\subsubsection{Scattering problem}

We start by solving the problem without tunneling not only between co-propagating modes but also at the QPCs.
The scattering problem involves four incoming and four outgoing channels: there are two incoming and two outgoing outer edge channels and two incoming and two outgoing inner edge channels. Fortunately, the chirality of the transport allows us to ignore  half of those channels, as we demonstrate below. This gives a major simplification in comparison with those QH states that have counter-propagating edge modes, such as the $\nu=2/3$ state.

Before we can proceed, we need to specify what sections of the inner and outer edge channels interact with each other in Fig.~\ref{fig:FP_interferometer}(a). 
We assume that the nearby inner and outer modes interact on the top and bottom sides of the device for all $x$ except in a narrow vicinity of points $x_1$ and $x_2$. No other other pairs of modes interact. In our model, the interaction strength is always the same whenever it is nonzero.

We next observe that nothing that happens downstream has any effect upstream (downstream and upstream mean down and up
with respect to the propagation direction of chiral modes). Thus, we can ignore the interaction of all outgoing channels on the left and on the right of the interferometer. More interestingly, we can ignore the interaction of the incoming modes. 
The reason is that the solution of the scattering problem below only requires knowledge of the time-dependent correlation functions of the outer incoming modes in point $x_1$ on the lower edge and point $x_2$ on the upper edge.
These correlation functions are the same as on an infinite edge that does not contain any inter-mode interactions downstream from the point $x_1$ for the bottom edge and $x_2$ for the top edge. Such an edge is in thermal equilibrium, and hence the time-dependent correlation functions in the point contacts are the same as if the inner channel did not exist at all. Thus, we can ignore the inner edge modes to the left of $x_1$ and to the right of $x_2$.

To understand the scattering process quantitatively, we will use the frequency-space representation of the Bose fields \cite{vondelft1998:bosonization}. 
For a left- or right-moving field with velocity $v$,%
\begin{equation}
	\phi(t,x) =	- \int_{0}^{\infty} \frac{d \omega}{\sqrt{\omega}}
	[ e^{-i\omega t}\phi(\omega, x) + \text{H.c.} ] e^{-\omega\delta/2},
\end{equation}
where
\begin{equation}
	\phi(\omega, x) = e^{\mp i \omega x/v} \phi(\omega, 0)   .
\end{equation}
$\phi(\omega,x)$ satisfies the following commutation relation,
\begin{equation}
	[ \phi(\omega, x), \phi^{\dagger} (\omega',x ) ] = \delta(\omega-\omega') .
\end{equation}
Such a relation applies in the limit of an infinitely long edge, which is relevant for the outer edge channels. 
The Bose fields that describe the closed inner channel in the interferometer drop out after the equations of motion are solved. The equations of motion are the first order differential equations for the charge densities $\sim\partial_x\phi$,
which are derivatives themselves.
We ignore that subtlety below and proceed as if one could remove one partial derivative with respect to the coordinates from all terms in the equations. This simplifies notations and does not affect our final results.

The outgoing fields depend on the incoming fields as follows,
\begin{equation}
	\begin{pmatrix}
		\phi_{\rm o}^{u}(\omega, x_{1})	\\
		\phi_{\rm o}^{d}(\omega,x_2)	
	\end{pmatrix}	= S
	\begin{pmatrix}
		\phi_{\rm o}^{d}(\omega,x_{1})	\\
		\phi_{\rm o}^{u}(\omega,x_{2})
	\end{pmatrix},
\end{equation}
with $S$ being an unitary $2\times2$ matrix.
Next, we try to find the matrix $S$ by solving the propagation equations and boundary conditions. A similar problem was solved in Ref.~\cite{ferraro2017:10.21468/SciPostPhys.3.2.014}, and we follow its approach here.
See also Ref.~\cite{isafi} for a related approach to quantum wires.
In the region between the two QPCs, due to the interaction, the free-propagating fields are the eigenmodes in Eq.~(\ref{eq:eigen_modes}), hence, they satisfy the following equations,
\begin{align}
	\phi_{j}^{d}(\omega, x_{2}) &= e^{i \omega a/v_{j}} \phi_{j}^{d}(\omega, x_{1}) ,\\
	\phi_{j}^{u}(\omega, x_{2}) &= e^{- i \omega a/v_{j}} \phi_{j}^{u}(\omega, x_{1}) ,
\end{align}
with $j=1,2$. In addition, the closed loop should satisfy the continuity conditions at the positions of the QPCs, where the interaction is minimal,
\begin{align}
	\phi_{\rm i}^{d}(\omega, x_{1}) &= \phi_{\rm i}^{u}(\omega, x_{1}), \\
	\phi_{\rm i}^{d}(\omega, x_{2}) &= \phi_{\rm i}^{u}(\omega, x_{2}). 
\end{align}
Solving for $\phi_{\rm o}^{u}(\omega, x_{1})$ and $\phi_{\rm o}^{d}(\omega,x_2)$, we can find
\begin{align}
	S_{11}& = -\frac{\sin ^2\theta \cos ^2\theta \left(e^{\frac{i a \omega }{v_{1}}}-e^{\frac{i a \omega }{v_{2}}}\right)^2}{-1+\left(\sin ^2\theta e^{\frac{i a \omega }{v_{1}}}+\cos ^2\theta e^{\frac{i a \omega }{v_{2}}}\right)^2}, \label{eq:S11}\\
	S_{12} & = \frac{2 i e^{\frac{i a \omega  (v_{1}+v_{2})}{v_{1} v_{2}}} \left[\sin ^2\theta \sin \left(\frac{a \omega }{v_{1}}\right)+\cos ^2\theta \sin \left(\frac{a \omega }{v_{2}}\right)\right]}{-1+\left(\sin ^2\theta e^{\frac{i a \omega }{v_{1}}}+\cos ^2\theta e^{\frac{i a \omega }{v_{2}}}\right)^2} .
	\label{eq:S12}
\end{align}
From the symmetry of the equations, $S_{11}=S_{22}$ and $S_{12}=S_{21}$.

\subsubsection{Correlation functions}

We can now compute $P_{\rm o}(t,x)$ at zero temperature using the formula for free Bose fields $A$, $B$, $C$ and $D$,
\begin{align}
	\langle{e^{iA} e^{-iB} e^{iC} e^{-iD}}\rangle = \exp\Big[ - \frac{1}{2} ( \langle A^2\rangle+\langle B^2\rangle+\langle C^2\rangle+\langle D^2\rangle ) \nonumber\\
	+ \langle AB\rangle -\langle AC\rangle +\langle AD\rangle +\langle BC\rangle -\langle BD\rangle +\langle CD\rangle  \Big],
\end{align}
and the zero-temperature correlation functions between free chiral fields $\phi$ and $\phi^\dagger$, 
\begin{align}
	\langle \phi(\omega, x) \phi^{\dagger}(\omega', x) \rangle &= \delta(\omega-\omega') \Theta(\omega), \\
	\langle \phi^{\dagger}(\omega, x) \phi(\omega', x) \rangle	&= 0%
\end{align}
(remember that $\omega$ is necessarily positive). All correlation functions can be expressed in terms of the correlation functions of the incoming outer edge modes at $x_1$ and $x_2$. The incoming fields in those two points are uncorrelated with each other due to chirality. Their self-correlation functions assume equilibrium values unaffected by the inner mode.

Computing $\langle \hat{T}(t,x_1) \hat{T}^{\dagger}(0, x_{1})\rangle$ gives
\begin{align}
	P_{\rm o}(t,0) 
	= \delta^{-2 m^2 \nu_1} \exp\Big\{ - m^2\nu_1 \int_{0}^{\infty} \frac{d\omega}{\omega} \nonumber\\
	  [ 2 - ( S_{11}+S_{11}^{*} ) ](1-e^{-i\omega t} ) e^{-\omega \delta}  \Big\},
\end{align}
and computing $\langle \hat{T}(t,x_2) \hat{T}^{\dagger}(0, x_{1})\rangle$ gives
\begin{align}\label{eq:P_outer}
	P_{\rm o}(t,a) 
	= \delta^{-2 m^2\nu_1} P_{1}(a) \exp\Big\{ - m^2\nu_1 \int_{0}^{\infty} \frac{d\omega}{\omega}  \nonumber\\
	[ 2  - ( S_{12}+S_{12}^{*} ) e^{-i\omega t} ] e^{-\omega \delta}  \Big\},
\end{align}
where
\begin{align}
	P_{1}(a) = \exp \Big( 2 m^2\nu_1 \int_{0}^{\infty} \frac{d\omega}{\omega} 
	S_{11}^{*} 	e^{-\omega \delta}  \Big)
\end{align}
is a real number. We note that $P_{\rm o}(t,a)$ satisfies the relation $P_{\rm o}(t,a) = P_{\rm o}(t,-a)$ as well.

We simplify the equations for the tunneling current as
\begin{align}\label{eq:current_non_int_outer}
	I_{\rm o, \text{non-int}} &= \frac{2 e^{*}}{\hbar^2} (|\Gamma_{1}|^2+|\Gamma_{2}|^2) 
	\int_{-\infty}^{\infty}dt \sin(\omega_{J} t) \Im P_{\rm o}(-t,0), \\ 
	\label{eq:current_int_outer}
	I_{\rm o, \text{int}} &= \frac{4 e^{*}}{\hbar^2} |\Gamma_{1} \Gamma_{2}| \cos\varphi \int_{-\infty}^{\infty} d{t} \sin{( \omega_{J} t)}  \Im P_{\rm o}(-t,a),
\end{align}
where $\exp(i\varphi)$ is the phase difference between $\Gamma_{1}$ and $\Gamma_{2}$. 

At a finite temperature $T$, we substitute the Bose-Einstein distribution into the correlation functions of the Bose fields, i.e.,
\begin{align}
	\langle \phi(\omega, x) \phi^{\dagger}(\omega', x) \rangle &= \frac{\delta{(\omega- \omega')}}{1-e^{-\hbar\omega/T}} , \\
	\langle \phi^{\dagger}(\omega, x) \phi(\omega', x) \rangle	&= \frac{\delta(\omega- \omega')}{e^{\hbar\omega/T}-1}.
\end{align}
Therefore we can find $P_{\rm o} (t,x)$ at a finite temperature as
\begin{widetext}
	\begin{align}
	P_{\rm o}(t,0) & = \delta^{-2 m^2 \nu_1} \exp\Bigg\{  - m^2\nu_1 \int_{0}^{\infty} \frac{d\omega}{\omega} 		
		[ 2 - ( S_{11}+S_{11}^{*} ) ] 
		\left(\frac{1-e^{-i\omega t}}{1-e^{-\hbar\omega/T}} + \frac{1-e^{i\omega t}}{e^{\hbar\omega/T}-1} \right)
		e^{-\omega \delta}  \Bigg\},\\
	P_{\rm o}(t,a) & = \delta^{-2 m^2 \nu_1} P_{1}(a)\exp \Bigg\{ - m^2 \nu_1  \int_{0}^{\infty} \frac{d\omega}{\omega}
		\Big[   
		\frac{2   - ( S_{12}+S_{12}^{*} ) e^{-i\omega t}}{1-e^{-\hbar\omega/T}}  +  
		\frac{2 - ( S_{12}+S_{12}^{*} ) e^{i\omega t}}{e^{\hbar\omega/T}-1} \Big] 
		e^{-\omega \delta} \Bigg\},
	\end{align}
\end{widetext}
where
\begin{align}
		P_{1}(a) = \exp\Big[ 2 m^2 \nu_1  \int_{0}^{\infty} \frac{d\omega}{\omega}  
	&\left(  \frac{S_{11}^{*}}{1-e^{-\hbar\omega/T}}
	+  \frac{S_{11}}{e^{\hbar\omega/T}-1}
	\right) \nonumber\\
	& e^{-\omega \delta}  \Big].
\end{align}

To numerically compute the current in Eq.~(\ref{eq:current_non_int_outer}) and (\ref{eq:current_int_outer}), one needs to compute the integral in the expression for $P_{\rm o}(t,x)$ first. 
Unfortunately, there is no known analytical result for $P_{\rm o}(t,x)$. However, following a technique similar to that used in Ref.~\cite{ferraro2017:10.21468/SciPostPhys.3.2.014}, one can write the integral in $P_{\rm o}(t,x)$ as a series. The details of this series expansion are explained in Appendix~\ref{appendix:series}.

\subsection{Aharonov-Bohm phase and the average current on the laboratory time scale}
\label{sec:II_E}

We now turn to the effect of voltage-induced fluctuations of the charge on the closed inner channel.
We compute the interference contribution to the current and represent it as 
\begin{equation}
    \label{basic}
    I_{\rm i, int}=\tilde I\cos\varphi,
\end{equation}
where $\varphi$ is an effective Aharonov-Bohm phase, accumulated by the interfering charges. 
Our focus above was on $\tilde I$. The rich physics of the phase $\varphi$ in the linear transport regime has been discussed in previous work on Coulomb-dominated interferometry \cite{halperin2011:PhysRevB.83.155440,review-FH}, and much of the same discussion applies to a well-screened case. It is not a goal of this manuscript to address the behavior of the phase in detail in the linear regime at finite temperatures. Some subtleties of its behavior are addressed in Appendices~\ref{appendix:additional_phase} and \ref{appendix:glitch}. In the main text, we limit ourselves with a few comments and focus instead 
on the limit of low temperatures with the emphasis on non-linear transport. This regime is greatly affected by intermode tunneling
between the inner and outer channels in the closed loop geometry of Fig.~\ref{fig:FP_interferometer}(b). 

The general structure of the phase factor is 
\begin{equation}
	\label{phase-nu-2}
	\exp(i\varphi)=\exp[ i C + i (\alpha_{0} \tilde{V} + \varphi_\lambda ) + i \gamma N ],
\end{equation}
where $C$ is a constant, depending on microscopic details, such as the structure of the tunneling contacts;
the $\alpha_0\tilde V$ term represents the dependence of the phase on the average voltage $\tilde{V}=(V_{d}+V_{u})/2$ due to the dependence of the area of the open channel on the chemical potential; 
and $\gamma$ represents the phase due to $N$ holes of positive charge $e$ confined on the inner channel at $\nu=2$ or due to $N$ confined quasiparticles of positive charge $e/3$ at $\nu=2/5$. In both cases, Coulomb interaction of the $N$ particles with the outer edge channel changes its area. At $\nu=2/5$, $\gamma$ also includes a statistical phase of anyons.
The value of $\gamma$ can always be chosen between $-\pi$ and $\pi$ 
without affecting the phase (\ref{phase-nu-2}) by adding a multiple of $2\pi$, and we will assume such a choice. 
Finally, $\varphi_\lambda$ represents an additional  phase shift that  depends linearly on any variations in control parameters such as gate voltages or the magnetic field. For simplicity, we shall focus on the case of a single control parameter, which we label $\lambda_1$, and we write $\varphi_\lambda = \alpha_1 \lambda_1$, where $\alpha_1$ is a constant.

The total charge on the inner channels is easy to compute when 
the QPCs are open. The charge  equals
\begin{equation}
    \label{intro-inner-charge}
    Q=|e^*| ( \tilde{U}+\beta_1\lambda_1 ),
\end{equation}
where  $e^*$ is the quasiparticle charge, $\tilde{U}$ is a linear function of $\tilde{V} $, and $\beta_1$ is another constant.
As discussed below, the charge of the closed inner edge is quantized. 
Still, Eq. (\ref{intro-inner-charge}) is an important reference point in that case, too. In particular, in the linear-response regime, $N=Q/|e^*|$ is the nearest integer to the prediction of the above equation.

\subsubsection{Linear regime} \label{jumps}

If tunneling happens between the inner edges of the device,  a simple model of a fixed-area interferometer may apply. The phase $\varphi$ then combines a statistical phase due to localized anyons in the bulk of the device (at $\nu=2/5$ only) and the standard Aharonov-Bohm phase proportional to the fixed area of the device, the magnetic field, and the charge of a tunneling quasiparticle. 

The situation is more complex when the inner channel is closed.
Indeed, in that case the fixed-area model does not work since the charge confined by the inner edge cannot change continuously in response to changing magnetic field or chemical potential. It is quantized as an integer number of electron charges at $\nu=2$ and an integer number of $e/3$ charges at $\nu=2/5$. 
This number can only change discontinuously as a new quasiparticle is added.
Since the interior of the sample is taken to be in an incompressible quantum Hall state, its electron density will increase with increasing magnetic field. As the enclosed charge is constant between jumps, the enclosed area will shrink, leaving a charge deficit in the vicinity of the edge.  The interior charge will be well screened by gates, but as the area of the inner mode shrinks, the induced charge on the inner edge will cause a change in phase of the outer edge due to the Coulomb coupling. In the limit of strong inter-mode coupling, the decrease in the charge of the inner edge will be precisely compensated by a charge increase on the outer edge, which is in good contact with the leads. Together with an equal charge increase resulting from the increased flux through the area of the outer mode, this will cause the interference phase to increase with magnetic field at twice the rate that one would observe in the absence of coupling between the modes. Thus, at $\nu=2$, with strongly coupled modes, the change in phase due to a small increase in magnetic field will be twice what would 
be observed at $\nu=1$ in the same magnetic field. This can indeed be seen experimentally
(see Ref.~\cite{Kim} and references therein).

From time to time, the charge within the inner channel will jump. This jump is screened by an opposite-sign jump of the charge within the outer channel. In the strong-coupling regime, the screening charge is exactly equal to one quasiparticle charge. Hence, at $\nu=2$, such jumps are invisible in interferometry.

Appendix~\ref{appendix:additional_phase} computes the dependence of $\varphi$ on the potentials $V_d$ and $V_u$ of the lower and upper edges in the intermediate coupling case. The dependence is linear between jumps of the charge of the inner channel. Such jumps do result in discontinuities of $\varphi$. The effects of these glitches on the Fourier transforms of the differential conductance with respect to various control parameters, which are the key quantities to be extracted from interferometer experiments, are analyzed in
Appendix~\ref{appendix:glitch}.

Glitches are only discontinuous at zero temperature. They are still relatively sharp, however, as long as the temperature is lower than the charging energy. 
At higher temperatures, the charge confined by the inner loop strongly fluctuates. The physics is similar to the fluctuations of the bulk charge in a poorly screened device. We do not address this effect here and refer the reader to earlier work \cite{halperin2011:PhysRevB.83.155440}.

The constraints on  the inner mode charge derived above can also be obtained from the Hamiltonian (\ref{HT}),  together with Eq.~(\ref {Top}), in the strong tunneling limit. Minimizing the energy then leads to the result 
\begin{equation}
n\sqrt{\nu_2} \frac {\Delta \phi_{\rm{i}}} {2 \pi} - n \nu_2 A_2\frac {(B-B_0)} {\Phi_0}  = 
{\rm{integer}},
\end{equation}
where $\Delta \phi_{\rm{i}}$ is the net change in $\phi_{\rm{i}}$ around the edges of the interferometer, $A_2$ is the area enclosed by the inner loop, $B_0$ is the reference magnetic field for which the ground state has $\partial_x \phi_{\rm{i}}=0$, $\Phi_0$ is the flux quantum, and the integer $n$ is the charge of the tunneling quasiparticle, divided by $-\nu_2 e $. The  charge on the inner edge is given by $e \sqrt{\nu_2} \Delta \phi_{\rm{i}}/2\pi $. For $\nu=2/5$, we have $\nu_2 = 1/15$ and $n=3$.

\subsubsection{Nonlinear regime}

We now address slow fluctuations of the charge confined by the closed inner edge due to the tunneling between co-propagating edge modes on each side of the interferometer in the closed-loop geometry.
We start with $\nu=2$.
\\

\paragraph{$\nu=2$ case.}

\begin{figure}[!htb]
	\centering
	\includegraphics[width=.7\columnwidth]{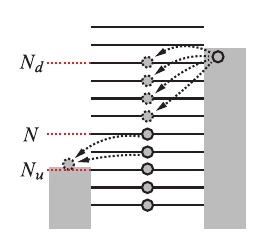}
	\caption{Schematics of the kinetics of quasiparticles on the inner loop in 
	the low-temperature limit. The dashed arrows on the right represent the tunneling 
	of quasiparticles from the lower edge into $N_{d}-N$ unfilled  states. The dashed 
	arrows on the left represent the tunneling  of quasiparticles from $N-N_{u}$ 
	filled  states into the upper edge.}
	\label{fig:levels}
\end{figure}

We will only address charge fluctuations in the low-temperature limit. This contrasts with our results in the short-time regime, which apply at any temperature. The fluctuations are driven by the potential difference between the upper and lower edges of the interferometer. Their origin can be visualized in the model of non-interacting electrons on the QHE edges at $\nu=2$, Fig. \ref{fig:levels}. The inner edge possesses a discrete set of electron energy levels. Electrons from the outer mode with a higher chemical potential, say at the lower edge, can tunnel into available states of the inner edge below the chemical potential. This is a slow process, involving spin flips. Between such rare tunneling events,  any neutral  excitations on the  inner edge equilibrate with the environment. The energy of the inner edge is not conserved, due to its Coulomb interaction with the outside world.  Thus, we expect the inner edge to relax into the lowest-energy state consistent with its charge. This happens on a much faster time scale than the tunneling events into the inner loop. At the same time, the chemical potential of the upper open edge is below the chemical potential of the inner loop. This leads to tunneling from the inner loop to the upper edge channel. Again, the inner loop rapidly sets into a ground state after any such tunneling event.

The above picture only applies in the non-linear transport regime. If the chemical potentials of the upper and lower edges are close to each other, the inner edge is unlikely to have an energy level between the two chemical potentials. This prevents the tunneling into the inner island.

In general, the tunneling rates between the inner loop and the outer edges depend on microscopic details. We will adopt a simple model in which the tunneling matrix elements are the same for all states of the inner edge. In other words, we simply assume that the tunneling rate between the lower and inner edges is proportional to the number of the empty energy levels on the inner edge below the chemical potential of the lower edge. 
The tunneling rate between the inner and upper edges is similarly proportional to the number of occupied states on the inner edge above the chemical potential of the upper edge. This describes Ohmic transport since the currents between the inner loop and outer edges are, on average, proportional to the chemical potential differences between the inner island and the outer edges. The quantization of the charge of the inner edge means that its zero-temperature chemical potential assumes a discrete set of values. The number of accessible values for a given voltage bias will depend on the magnetic flux or other parameters, represented by the control parameter $\lambda_1$ and changes by $\pm 1$  as $\lambda_1$ is varied. In general, at a fixed bias, two different level numbers are possible, say $k$ and $k+1$, depending on  $\lambda_1$.
As the bias is increased, one will cross a threshold where the number of allowed levels changes to $k+1$ and $k+2$.  Quantities such as the Fourier amplitudes of the current with respect to 
the magnetic field or other control parameters will exhibit mathematical singularities as these thresholds are crossed.

The range of possible hole numbers $N$ on the inner edge as a function of $V_d$ and $V_u$ is computed in Appendix \ref{appendix:additional_phase}. 
We shall denote the minimal and maximal allowed values for given $V_d$ and $V_u$ and fixed $\lambda_1$   as $N_u<N_d$, assuming $V_d > V_u$.

We now introduce a kinetic equation for $N$.
The distribution function of the number of holes $N$ is $f_{N}$, and we consider its rate of change. When there are $N-1$ holes on the inner edge, the holes on the lower edge have enough energy to tunnel into any one of the unfilled $N_{d}-(N-1)$ levels, and we assume that the tunneling rates are all equal to $\Gamma_{\rm io}$. 
The rate of the $N-1\rightarrow N$ transition is $\Gamma_{\rm io} f_{N-1} [ N_{d}-(N-1) ]$. Similarly, the rate of the $N+1\rightarrow N$ transitions is $\Gamma_{\rm io} f_{N+1} (N+1-N_{u}) $. Finally, the combined rate of the $N\rightarrow N-1$ and $N\rightarrow N+1$ processes is $\Gamma_{\rm io} f_{N} [(N-N_{u})+(N_{d}-N)]$. Therefore, we have the following kinetic equation:
\begin{align}
	\dot{f}_{N}= &- \Gamma_{\rm io} f_{N} (N_{d}-N_{u}) + \Gamma_{\rm io} f_{N-1} [ N_{d} - (N-1) ] \nonumber\\
	&+ \Gamma_{\rm io} f_{N+1} (N+1-N_{u}).
\end{align}
The stationary distribution function for $N$ can be found to be
\begin{equation}
	f_{N} = \frac{C_{N_{d}-N_{u}}^{N-N_{u}}}{2^{N_{d}-N_{u}}}.
\end{equation}

We will now average the interference current over $f_N$. This requires us to 
average the phase factor (\ref{phase-nu-2}) in Eq.~(\ref{basic}), where we use 
$\alpha_0=-{2ea}/{\hbar v_{\rm o}}$, Eq.~(\ref{eq_alpha_0}), and
$\gamma= {2\pi w}/{v_{\rm o}} \pmod{2\pi}$, $-\pi <\gamma\leq \pi$, Eq.~(\ref{eq_gamma}).

We observe that from the binomial formula,
\begin{align}
	\frac{1}{2^{N_{\rm max}}} \sum_{N=0}^{N_{\rm max}} C_{N_{\rm max}}^{N} e^{i \theta N} 
	&= \frac{1}{2^{N_{\rm max}}} (1+ e^{i\theta})^{N_{\rm max}}  \nonumber \\
	&= (e^{i\theta/2})^{N_{\rm max}} \cos^{N_{\rm max}}(\theta/2) .
\end{align}
Hence, the expected value of the phase factor is
\begin{align}
	\label{average-phase-nu-2}
	&  \sum_{N=N_{u}}^{N_{d}} f_{N} \exp[ i C + i (\alpha_{0} \tilde{V} + \alpha_{1} \lambda_{1} ) + i \gamma N ] \nonumber\\
	= & \exp\Big\{  i C + i (\alpha_{0} \tilde{V} + \alpha_{1} \lambda_{1} ) + i \frac{\gamma}{2}( g_{u}+g_{d} )\nonumber\\ 
	& + (2\beta_{0} \frac{v_{\rm o}}{w}) \log(\cos{\frac{\gamma}{2}}) V
	 + F(g_{d}) \left[ -i\frac{\gamma}{2} -\log(\cos{\frac{\gamma}{2}}) \right]\nonumber\\
	& + F(g_{u}) \left[-i \frac{\gamma}{2} + \log(\cos{\frac{\gamma}{2}}) \right] \Big\} ,
\end{align}
where $V=V_{d}-V_{u}$,
and we use the function $F(g) = g-[g]$, where $[x]$ is the nearest integer to $x$, 
and $g_u$ and $g_d$ are computed in Appendix~\ref{appendix:additional_phase} as
\begin{align}
	\label{eq:beta_1_u}
	g_{u} &= \beta_{0} \left( \frac{2 v_{\rm o}}{w} V_{u} - V_{u} - V_{d}  \right) + \beta_{1} \lambda_{1}; \\
	\label{eq:beta_1_d}
	g_{d} &= \beta_{0} \left( \frac{2 v_{\rm o}}{w} V_{d} - V_{u} - V_{d} \right) + \beta_{1} \lambda_{1}.
\end{align}
As discussed above [Eq. (\ref{intro-inner-charge})] and in Appendix~\ref{appendix:glitch}, $\lambda_1$  represents a selected control parameter, such as a gate voltage or the magnetic flux through the inner loop, and $\beta_1$ is a constant that depends on microscopic details.

A striking feature of the above expression is the $(V_d-V_u)(\log\cos\frac{\gamma}{2})$ term, responsible for the exponential suppression of the current at a large voltage bias $V_d-V_u$.
Note also that if $\gamma=\pi$, there is no interference current 
when $N_u\ne N_d$.
\\

\paragraph{$\nu=2/5$ case.}
We will use an Ohmic model similar to the $\nu=2$ problem. This may seem unjustified since the tunneling of fractionally charged quasiparticles is typically non-Ohmic in the FQHE. 
That is, however, not the case for tunneling between co-propagating edge channels. 
Indeed, the edge theory with co-propagating modes can be interpreted as a chiral conformal field theory. Tunneling terms in the Hamiltonian must be Bose fields, as is the case for any contribution to the Hamiltonian. Hence, their scaling dimension is integer.
That integer is usually 1. In particular, this is the case for the inter-channel tunneling of $e/3$ anyons at $\nu=2/5$. Since the scaling dimension is the same as at $\nu=2$, the use of a similar model is justified.

We again find a binomial distribution $f_N$ for the quasiparticle number. 
The statistical 
phase is computed in Appendix \ref{appendix:additional_phase} and equals 
${\gamma} = 2\pi \nu_{1} ( \frac{\sqrt{\nu_{1}}w}{\sqrt{\nu_{2}} v_{\rm o}} - 1)$. We can always 
assume that $-\pi<{\gamma}\leq\pi$ by adding an irrelevant multiple of $2\pi$. We 
also compute the coefficient $\alpha_{0}=-2\nu_{1}ea/\hbar v_{\rm o}$ in Appendix~\ref{appendix:additional_phase}. 
Finally, we average the phase factor over $f_N$ and find
\begin{align}
\label{eq:53}
	&\sum_{N=N_{u}}^{N_{d}} f_{N} \exp[ i C + i (\alpha_{0} \tilde{V} + \alpha_{1} \lambda_{1} ) + i \gamma N ]\nonumber\\
	=& \exp\Big[ \begin{aligned}[t]
		&i C + i (\alpha_{0} \tilde{V} + \alpha_{1} \lambda_{1} ) + i \frac{\gamma}{2}(N_{d}+N_{u}) 
	\nonumber\\ &+ (N_{d}-N_{u}) \log(\cos{\frac{\gamma}{2}})\Big] \nonumber\\
	\end{aligned}\\
	=& \exp\Big\{ i C + i (\alpha_{0} \tilde{V} + \alpha_{1} \lambda_{1} ) + i \frac{\gamma}{2}( g_{u}+g_{d} )\nonumber\\ 
	&+ (2\beta_{0} \frac{v_{\rm o}}{w}) \frac{\nu_{2}}{\nu_{1}} \log(\cos{\frac{\gamma}{2}}) V 
	+ F(g_{d}) \left[ -i\frac{\gamma}{2} -\log(\cos{\frac{\gamma}{2}}) \right] \nonumber\\
	&+ F(g_{u}) \left[-i \frac{\gamma}{2} + \log(\cos{\frac{\gamma}{2}}) \right] \Big\},
\end{align}
where the definitions of $g_u$ and $g_d$ have been modified according to the discussion of $\nu=2/5$ in Appendix \ref{appendix:additional_phase}:
\begin{align}
	g_{u} &= \frac{\sqrt{\nu_{2}}}{\nu_{1}}\beta_{0} \left( \frac{2 v_{\rm o}}{w}  \sqrt{\nu_{2}} V_{u} - \sqrt{\nu_{1}} V_{u} - \sqrt{\nu_{1}}V_{d}  \right) +  \beta_{1} \lambda_{1} \\
	g_{d} &= \frac{\sqrt{\nu_{2}}}{\nu_{1}}\beta_{0} \left( \frac{2 v_{\rm o}}{w} \sqrt{\nu_{2}} V_{d} - \sqrt{\nu_{1}} V_{u} - \sqrt{\nu_{1}} V_{d} \right) +  \beta_{1} \lambda_{1}.
\end{align}

As at $\nu=2$, the above expression reveals exponential suppression of the current at a high bias $V=V_u-V_d$. As before, $\gamma=\pi$ implies no interference current for $N_u\ne N_d$.

\subsection{Asymmetric bias} \label{sec:filling_two_asymmetric}

In the previous subsections, we have discussed the response under the assumption that the bias voltage $V$ is applied symmetrically to the two edges, i.e., that the voltages $V_d$ and $V_u$ applied to the lower and upper edges are equal to $V/2$ and $-V/2$ respectively. However in many experiments, the bias is applied asymmetrically, and in many cases only to one of the edges.  This is the case, for example, in the experiments of Ref.~\cite{Kim}.   In the general case we may write 
\begin{equation}
	V_{d/u} = \tilde  {V} \pm V/2
\end{equation}
where $\tilde {V}$ is the average of the applied voltages and $V$ is their difference. 
A non-zero value of $\tilde{V}$ can introduce an additional phase factor into the tunneling amplitudes in the presence of screening gates~\cite{law2006:PhysRevB.74.045319}. This reflects the charge, accumulated in the interferometer due to the applied voltage. Indeed, the phase accumulated by a particle on a path around an interferometer is proportional to the charge inside the device~\cite{fh2022}.

In this subsection we address the effect of non-zero $\tilde V$. We limit our discussion to $\nu=2$ since the relevant experiments \cite{nakamura2023:fabry} at $\nu=2/5$ involve symmetric bias. 
We  also assume a fixed charge of the inner island, since the fluctuations of the island charge
do not differ qualitatively between the cases of symmetric and asymmetric bias.

If $\tilde{V} \neq 0$ but  $V=0$, the system will remain in a state of thermal equilibrium, with some alteration of the total charge, which will generally lead to a shift in the interference phase $\varphi$.
At a low temperature, the charge of the inner channel assumes a fixed value that minimizes the energy shifted by the total charge times the chemical potential. In what follows in this subsection, we shall assume the inner loop charge is fixed at this value, even at a nonzero $V$.   

Application of an asymmetric bias
to the device will then lead to an interference current with an altered phase but with an amplitude computed in precisely the same way as in the case of a symmetric bias.  Thus we may write
\begin{equation}
	I_{ \text{int}} =	\tilde {I} \cos\varphi ,
\end{equation}
where  $\varphi$  depends on $\tilde{V}$ but is independent of $V$, while 
$\tilde {I}$   depends  on   $V$ but not on $\tilde{V}$.

An important experimental quantity is differential conductivity, $dI/dV$.  
If the voltage is applied asymmetrically, with $\tilde{V} = \eta V$, we have
\begin{equation}
	\label{dIdV}
	\frac {d  I _{\text{int}}} {dV} = \frac{ d \tilde{I} }{dV} \cos (\varphi)  
	-  \eta \tilde{I} \sin (\varphi) \frac {d \varphi} {d\tilde{V}}  .
\end{equation}
Typically, values of $dI/dV$ are obtained over a range of control parameters, such as the magnetic field and/or one or more gate voltages, and one takes a Fourier transform of the data with respect to these parameters. The transform will have peaks at frequencies corresponding to the interference oscillations, and we wish to compute the amplitudes of these peaks.
Asymmetry in the applied voltage can have an important effect on these amplitudes. 

In the case of a Fabry-P\'erot interferometer at $\nu=1$, the factor $d \varphi /d \tilde {V} $ is a constant,  equal to $-2ea/\hbar v$
(see Ref.~\cite{law2006:PhysRevB.74.045319}).
However, the case of $\nu=2$ is more complicated. In particular, for interference of the outer mode, when the inner mode is completely reflected at the point contacts, the charge on the inner edge is restricted to integer values. 
At low  temperatures, for $V=0$, the accumulated charge on the inner edge minimizes the energy of the system. Since only an integer number of electrons can tunnel into the inner edge, the accumulated charge changes discontinuously as one more electron is added or removed in response to the changing $\tilde {V}$. This may result in a discontinuous jump in $\varphi$  due to the Coulomb interaction of the inner and outer edge modes. As a result, the interference phase will not depend linearly on $\tilde{V}$ or on parameters such as $B$ or the 
gate voltages, and the interference current is no longer a simple sinusoidal function of these parameters. 
As discussed in Appendix~\ref{appendix:glitch}  this situation leads to an array of Fourier peaks, whose amplitudes may be analyzed separately. In the remainder of this subsection, we shall ignore this complication and assume that   $\varphi$ is a linear function of $\tilde{V}$ and the other parameters,  so that $ d \varphi / d \tilde{V} $ is a constant. However, as we  show in Appendix~\ref{appendix:glitch},  the essential results for the amplitude of the dominant finite-frequency Fourier peaks are unchanged after the phase  jumps are properly taken into account, except for renormalization by a voltage-independent constant.

This linearity assumption is actually  valid in the limit of very strong mode coupling, where  an integer  jump in the charge of the inner  mode will be compensated by an opposite integer jump in the charge of the outer mode. The interference phase will consequently jump by a multiple of $2 \pi$, which will have no effect on the  interference signal.  In this limit, therefore, it is appropriate to ignore the jumps, so that $\varphi$ is   a linear function of $\tilde{V}$. This near-invisibility of phase jumps in the strong-coupling regime is also responsible for the appearance of a flux period normally associated with  particles of charge $2e$ (cf. Appendix~\ref{appendix:glitch}).    In this case, we find 
\begin{equation}
	\label{dphidV}
	\frac {d \varphi }{d \tilde {V} } = {-\frac {2ea} {\hbar v_{\rm o}} },
\end{equation}
as  shown in Appendix \ref{appendix:additional_phase}.

In the discussions above, we ignored any effects on the edge modes due to changes in the occupation of localized states in the bulk of the interferometer. This agrees with our assumption that  the quantum Hall system is well screened by nearby gates.
If this is not the case, the slope $|{d \varphi }/{d \tilde {V} }|$ may be reduced below the value $2ea/ \hbar v_{\rm o} $.

From the analysis of the previous section, we obtain for the outer mode
\begin{align}
	\tilde{I} &= -\frac{4e}{\hbar^2} |\Gamma_{1} \Gamma_{2}|\int_{-\infty}^{\infty} d{t} \sin( \omega_{J} t ) \Im P_{\rm o}(-t,a).
\end{align}
Now, if we perform the Fourier transform of (\ref{dIdV}) with respect to $\varphi$, the amplitude $\mathcal{A}$ is
\begin{equation}
	\mathcal{A} = \left[\left(\frac{d \tilde{I} }{dV } \right)^2 + \left(\eta \frac {d \varphi }{d \tilde {V} } \tilde{I}\right)^2\right]^{\frac{1}{2}}.
	\label{eq:Fourier_amplitude}
\end{equation}

\section{Results for $\nu=2$ QH liquids}
\label{sec:filling_two}

In this section, we discuss our numerical results for $\nu=2$ QH liquids, where $\nu_1=\nu_2=1$. %
We first consider an open inner mode and then address a harder and more interesting problem of a closed inner island.
In subsection \ref{sec:III_A}, we consider
the case where the charge, confined by the inner loop, is time-independent.
We first consider the easier problem of symmetrically applied bias. We next address asymmetric bias.
We discuss non-linear transport in
the regime of fluctuating charge on the inner island in subsection \ref{sec:III_C}. 
Note that the results of subsections \ref{sec:III_A} 
apply  in the linear transport regime, even if the tunneling amplitude between co-propagating modes is not negligible, since at low temperatures, in the linear regime, the charge of the inner loop does not fluctuate. 
We assume an approximate symmetry between the spin-up and -down channels so that
$v_{\rm o}\approx v_{\rm i}$ and $\theta=\pi/4$.

\subsection{Inner mode tunneling}
\begin{figure}[htbp]
	\centering
	\includegraphics[width=.8\columnwidth]{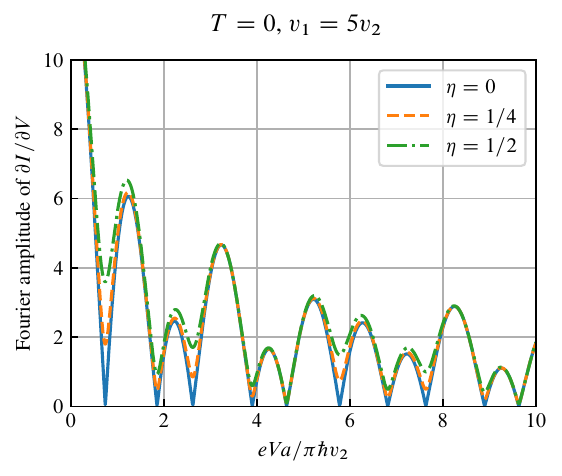}
	\caption{Inner mode tunneling at $\nu=2$. Fourier amplitudes $\mathcal{A}$ of the differential conductance $\partial I/\partial V$ with different choices of $\eta$ are plotted. The ratio between the faster and slower velocities of the normal modes is fixed at $v_{1}=5 v_{2}$, and the temperature $T=0$.}
	\label{fig:nu_2_inner}
\end{figure}

We first briefly discuss the inner mode tunneling case for $\nu=2$ interferometers. In this case, the bulk 
would not form a closed island, and the linearity of $\varphi$ over $\tilde{V}$ always holds, 
so that
\begin{equation}
	\frac {d \varphi }{d \tilde {V} } = {-\frac {2ea (v_{\rm o}-w)} {\hbar (v_{\rm i}v_{\rm o}-w^2)} }
	\approx - \frac{2ea}{\hbar v_{1}},
\end{equation}
where the second equality only holds at $\theta=\pi/4$.
To derive this equation one needs to remember that the same bias is applied to all open channels.
The Fourier amplitude 
of the interference contribution to the conductance
is thus given by Eq.~(\ref{eq:Fourier_amplitude}).
We have plotted the Fourier amplitude curves for different choices of asymmetry factor $\eta$ in Fig.~\ref{fig:nu_2_inner}.  One sees that asymmetrically applied voltage bias lifts the curve. 

\subsection{Outer mode tunneling: fixed island charge}
\label{sec:III_A}

Here we ignore tunneling between the inner and outer edge channels and hence assume a fixed charge 
of the inner closed loop. We start with the simplest limit of the symmetric voltage bias. In that case
$\tilde V=0$ in Eq. (\ref{phase-nu-2}). The Fourier transform of the current with respect to the control parameter $\lambda_1$, Eq. (\ref{phase-nu-2}), is trivial and essentially reduces to computing $\tilde I$.

Since $v_{\rm o}\approx v_{\rm i}$ due to the approximate symmetry of the spin-up and -down channels, the edge is in the strong interaction regime, i.e., $\theta=\pi/4$. The interaction strength $w$ is reflected in the normal-mode velocities $v_{1,2}$, where $v_1$ is the velocity of the charged mode and $v_2$ is the much slower velocity of the neutral mode. 
A related problem was considered in Ref.~\cite{ferraro2017:10.21468/SciPostPhys.3.2.014} in the limit of an infinite velocity of the charged mode. This limit is problematic from the point of view of locality since all events are causally related in a system with an infinite velocity of excitations.  
Ref.~\cite{ferraro2017:10.21468/SciPostPhys.3.2.014} focuses on zero temperature only and contains no plots of the voltage dependence of the interference current. Such plots are the main focus of this section, where we also address finite temperatures. In subsection~\ref{sec:Asymmetric_2} we also address the asymmetry of the voltage bias.

We shall concentrate on the Fourier amplitude      of $dI/dV$.   

\subsubsection {Symmetric bias}

\begin{figure}[!tbp]
	\centering	\includegraphics[width=.8\columnwidth]{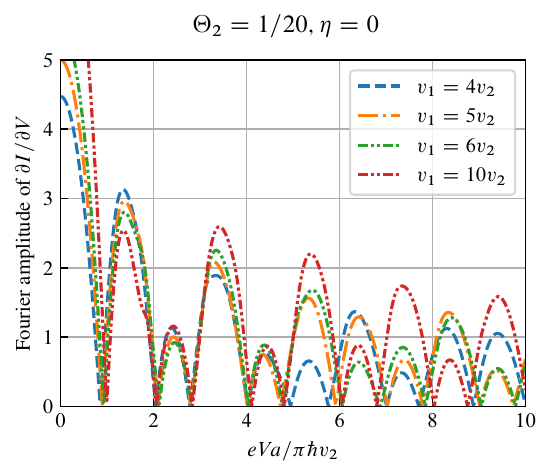}
	\caption{Outer mode tunneling at $\nu=2$ at a fixed island charge. The plots show the dependence of the Fourier amplitude %
	of the interference contribution to $\partial I/\partial V$ on the voltage bias, for different velocity ratios, when the bias voltage is applied symmetrically. The temperature is represented with a dimensionless quantity $\Theta_2 = T a/\hbar v_2$, and is fixed at $\Theta_2=1/20$.  %
	}
	\label{ref:Fig-2-sym}
\end{figure}

The results of calculations along the lines of Section \ref{sec:II_D} and Appendix \ref{appendix:series} are illustrated in Fig. \ref{ref:Fig-2-sym}, for the case of symmetric bias. 
In contrast to the $\nu=1$ interferometer at a symmetric bias (see Ref.~\cite{chamon1997:PhysRevB.55.2331}), where 
\begin{equation}
	\label{49}
	\tilde{I} = \frac{e}{\hbar^3} (\Gamma_{1} \Gamma_{2}^{*} + \Gamma_{1}^{*} \Gamma_{2}) \frac{4\pi^2 T}{\sinh(2\pi T a/\hbar v )} \sin(\omega_{J} a/v),
\end{equation}
these curves are no longer periodic in the voltage.
Despite the absence of any periodicity,
note a series of nodes
with the distance of roughly $\pi$ in units of $eV a/\hbar v_2$. This corresponds to $eV=2\pi\hbar v_2/2a$, where $2a$ is the perimeter of the interferometer, that is, to the lowest excitation energy on the loop of length $2a$ ($v_2$ is considerably slower than $v_1$).

Experiments in graphene were performed at a highly asymmetric voltage bias. This modifies the $I$--$V$ curves as discussed in Section~\ref{sec:filling_two_asymmetric} and below.

\subsubsection{Asymmetric bias}

\label{sec:Asymmetric_2}
\begin{figure}[!tbp]
	\centering
	\subfloat[]{\includegraphics[width=.8\columnwidth]{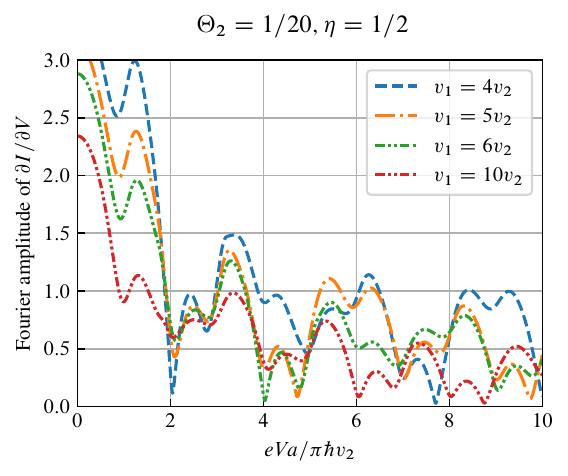}}\\
	\subfloat[]{\includegraphics[width=.8\columnwidth]{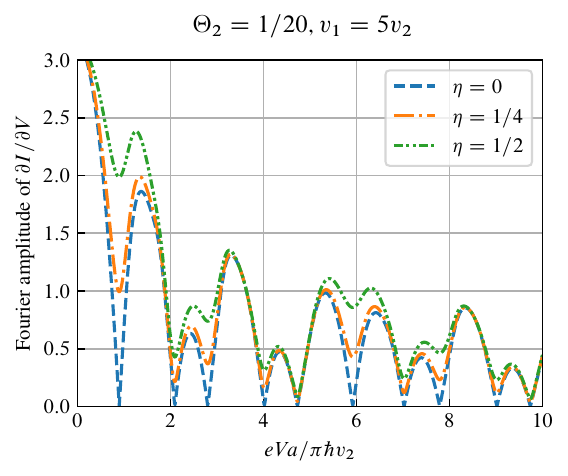}}
	\caption{Outer mode tunneling at $\nu=2$ at a fixed island charge. The plots show the dependence of the Fourier amplitude $\mathcal{A}$, Eq. (\ref{eq:Fourier_amplitude}), on the voltage bias at a finite temperature,  when the bias voltage is applied asymmetrically. The temperature is represented with a dimensionless quantity $\Theta_2 = T a/\hbar v_2$, and is fixed at $\Theta_2=1/20$. 
	(a) Fourier amplitude curves at different velocity ratios. 
	(b) The ratio between the faster and slower velocities of the normal modes is fixed at $v_{1}=5 v_{2}$, and Fourier amplitude curves with different choices of $\eta$ are plotted. 
	Nonlinearities in the dependence of $\varphi$ on $\tilde{V}$ and the control parameters have been ignored.
    }
	\label{fig:nu_2_Fourier}
\end{figure}

As above, a linear dependence of the interference phase on the control parameters is assumed.
Several plots of the Fourier amplitude at finite temperature are shown in Fig.~\ref{fig:nu_2_Fourier}. 
When the asymmetry $\eta$ and bias voltage $V$ are both small, the Fourier amplitude is close to $d \tilde{I}/dV$, which represents the differential conductance at $\varphi_{\rm AB}=0$.
When $\eta$ is greater, the values of the  Fourier amplitude at the nodes increase and are non-zero, but the positions of the first few minima are relatively unchanged.

Even though no periodicity is seen, the plots exhibit a series of nodes
whose distance is roughly $\pi$ in units of $eV a/\hbar v_2$ just like in the symmetric case. This corresponds to $eV=2\pi\hbar v_2/2a$, where $2a$ is the perimeter of the interferometer. This is the lowest excitation energy on the loop of length $2a$, as we already saw above. The plots show the absolute value of the Fourier amplitude. It is thus natural to associate the effective period with twice the  distance between successive nodes. We then discover approximately the same period as in Eq. (\ref{49}) at $v=v_2$. 

The above results apply in the weak backscattering regime. A perturbative calculation is insufficient for stronger tunneling. Note that for Fabry-P\'erot interferometers, an exact solution is known at $\nu=1$~\cite{chamon1997:PhysRevB.55.2331}. This solution only works in the absence of inter-channel interaction near QPCs.

\subsection{Outer mode tunneling: fluctuating island charge}
\label{sec:III_C}

We present here results for the interference current, averaged over  fluctuations in the island charge. Since we only solved the kinetic problem at zero temperature, the results apply only at low temperatures. At the same time, we consider an arbitrary voltage bias.

The current follows equation (\ref{basic}) with the averaged phase factor (\ref{average-phase-nu-2}).
Motivated by the usual way to present experimental data, we focus on the Fourier transform of $dI/dV$ with respect to a control parameter $\lambda_1$ such as the magnetic field. Technical details of the Fourier transform can be found in Appendix \ref{appendix:Fourier}. 

\begin{figure}[!tbp]
	\centering
	\subfloat[]{\includegraphics[width=.8\columnwidth]{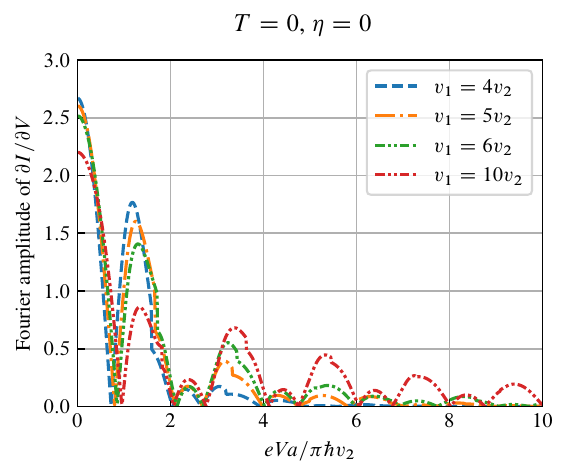}}\\
	\subfloat[]{\includegraphics[width=.8\columnwidth]{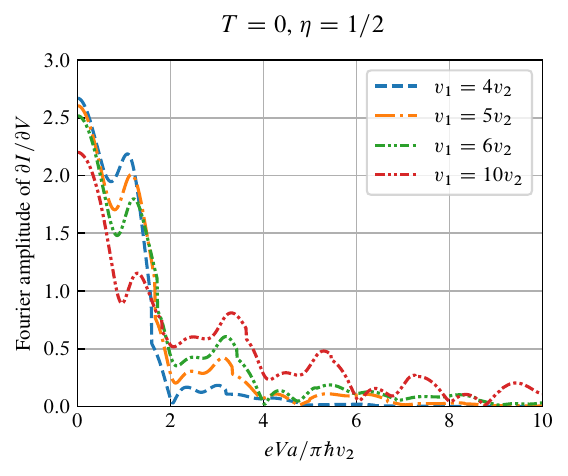}}\\
	\subfloat[]{\includegraphics[width=.8\columnwidth]{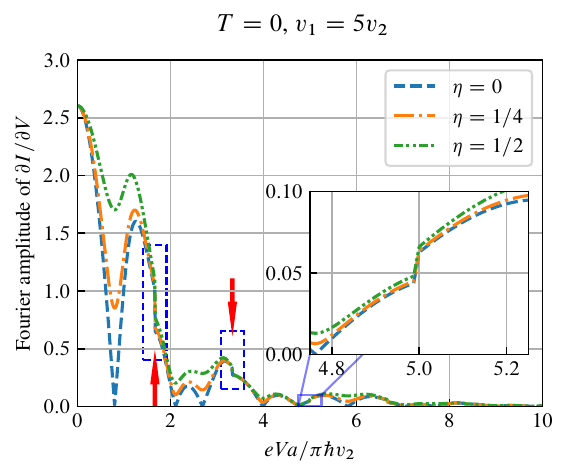}}
	\caption{Outer mode tunneling in $\nu=2$ interferometers for fluctuating island charge. (a) The dependence of the dominant Fourier amplitude on the voltage bias at zero temperature with a symmetric bias. (b) Fourier amplitude curves at different velocity ratios with a totally asymmetric bias. (c) The ratio between the faster and slower velocities of the normal modes is fixed at $v_{1}=5 v_{2}$, and Fourier amplitude curves with different choices of $\eta$ are plotted.
    Rectangular boxes and the inset show threshold features due fluctuations of the inner island charge that arise from residual tunneling between co-propagating modes.
	}
    \label{fig:nu_2_long}
\end{figure}

We now show results for several choices of the parameters in Fig.~\ref{fig:nu_2_long}. 
We plot the voltage dependence of one prominent Fourier harmonic with respect to $\lambda_1$.
In all cases we choose the harmonic of frequency $\alpha_1+{\gamma\beta_1}$, Eq. (\ref{average-phase-nu-2}). This is a natural choice since $\alpha_1\lambda_1$ should be understood as a naive Aharonov-Bohm phase $2\pi\Phi/\Phi_0$, where $\Phi$ is the magnetic flux through the nominal area of the device, and $\Phi_0$ is a flux quantum, and we can interpret $\gamma\lambda_1\beta_1$ as the phase due to the average $\lambda_1$-dependent charge on the inner island. This choice also corresponds to $m=0$ in Appendix \ref{appendix:Fourier}. As discussed in that Appendix, dominant harmonics correspond to small $m$. They are also likely less sensitive to temperature effects.
The value of $\beta_1$ (\ref{eq:beta_1_u},\ref{eq:beta_1_d}) does not matter for the plots because it only affects the scale of the vertical axis, which is arbitrary here. In all cases we set $T=0$. The rest of the parameters are listed in the figures. The key difference from  the case of the fixed island charge consists in a rapid decay of the signal at large $V$.

Another interesting feature is present at a discrete set of voltages. See, in particular, rectangular boxes and the inset in Fig. \ref{fig:nu_2_long}(c). One observes discontinuities in the Fourier amplitude of  $dI/dV$. Their origin is due to the tunneling between the outer modes and the inner island.
As shown in Appendix \ref{appendix:additional_phase}, the maximal and minimal 
hole numbers $N_d$ and $N_u$ on the inner loop equal $N_{d,u}=[g_{d,u}]$, where $g_{d,u}$ are given by Eqs. (\ref{eq:gu},\ref{eq:gd}) and the square brackets denote the nearest integer. If $g_d-g_u=2\beta_0 v_{\rm o}(V_d-V_u)/w$ is an integer, then the number $N_d-N_u+1$ of the energy levels, available for transport between the outer edges through the inner loop, is exactly $g_d-g_u+1$ irrespective of the flux $\lambda_1$ (see Appendix \ref{appendix:additional_phase}). For non-integer $g_d-g_u$, the number of the available levels depends on the flux and varies between the largest integer less than $g_d-g_u+1$ and the lowest integer greater than $g_d-g_u+1$. In other words, the structure of the set of available levels as a function of the flux changes at the integer values of $g_d-g_u$. This results in singularities in the Fourier transform with respect to $\lambda_1$ at
\begin{equation}
    \label{eq:V_nu_2}
    V=\frac{mw}{2\beta_0 v_{\rm o}}
\end{equation}
at each integer $m$, with $\beta_0$ given by (\ref {beta0}).  This feature can be used to extract the parameters of the edge theory from the data.

\section{Results for $\nu=2/5$ FQH liquids}\label{sec:filling_two_fifth}

In this section, we consider the interferometer in Fig.~\ref{fig:FP_interferometer} for $\nu=2/5$ QH liquids, where $\nu_1=1/3$ and $\nu_2=1/15$. We only study the case of symmetric voltage bias since the bias was applied symmetrically in the relevant experiment \cite{nakamura2023:fabry}.

\subsection{Inner mode tunneling}
\begin{figure}[!htb]
	\centering
	\includegraphics[width=.8\columnwidth]{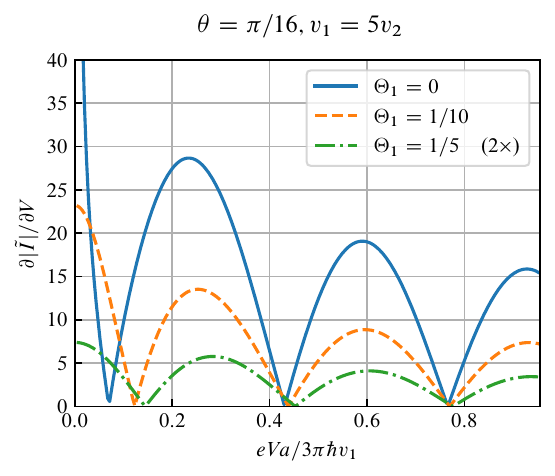}
	\caption{Inner mode tunneling at $\nu=2/5$. The dependence of the Fourier amplitude of the differential conductance of the interference current on the voltage bias is shown at different temperatures, with $\theta=\pi/16$ and $v_{1}=5v_{2}$. $\Theta_{1}$ is the dimensionless temperature $T a/\hbar v_1$.
	}
	\label{fig:inner_numerical_results:temperature}
\end{figure}

\begin{figure}[!htb]
	\centering
	\includegraphics[width=.8\columnwidth]{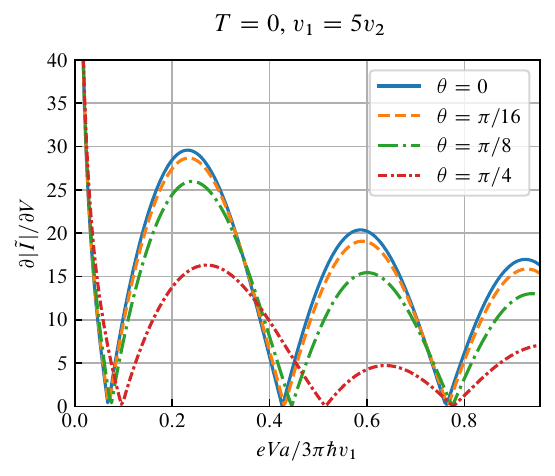}
	\caption{Inner mode tunneling at $\nu=2/5$. The dependence of the Fourier amplitude of the differential conductance on the voltage bias is shown at different interaction strengths (represented with $\theta$), with $v_1 =5 v_2$ and $T=0$.}
	\label{fig:inner_numerical_results:theta}
\end{figure}

For the inner mode tunneling, we numerically computed the differential conductance of the interference current, and the results are shown in Figs.~\ref{fig:inner_numerical_results:temperature} and \ref{fig:inner_numerical_results:theta}. In Fig.~\ref{fig:inner_numerical_results:temperature}, we plot the voltage dependence of the Fourier amplitude of the interference contribution to the conductance with respect to the magnetic flux at a weak coupling $\theta=\pi/16$. One can observe that the location of the first node depends significantly on the temperature, while the locations of the subsequent nodes show little temperature dependence. 
Nevertheless, the first node position remains finite at $T=0$. 
This is in contrast with the situation at $\nu=1/3$, 
where the first node of $dI/dV$ does move to zero voltage for $T \to 0$. The difference in behavior is related to the fact that 
at $\nu=1/3$, the tunneling current across a constriction follows $I\sim V^{2\times 1/3 -1}=V^{-1/3}$ at $T=0$, so  
the current grows at low voltages, so $dI/dV < 0.$ When $T \neq 0$, however, the system enters a linear regime, for $eV < T$, where $dI/dV >0$. Hence there is a sign change at a voltage of order $T/e$. By contast, at
$\nu=2/5$,  for inner-edge tunneling, the low-voltage current follows the relation
$I\sim V^{2\times 3/5-1}=V^{1/5}$  at $T=0$,   which goes to zero when $V \to 0$.  (The precise forms only apply as long as the conductance
$G\sim V^{-4/5}$ remains small.)


\subsection{Outer mode tunneling}\label{sec:two_fifth_outer}

\subsubsection{Fixed island charge}
\label{sec:IV_B_1}

In this subsection, we assume that 
charge of the inner closed loop is fixed at a value independent of the voltage bias and the magnetic flux. We also assume a symmetric application of the voltage bias. It is then straightforward to plot the Fourier transform of the interference contribution to the conductance with respect to the magnetic flux.

\begin{figure}[!htb]
	\centering
	\subfloat[\label{subfig:outer_numerical_results:temperature_a}]{%
		\includegraphics[width=.8\columnwidth]{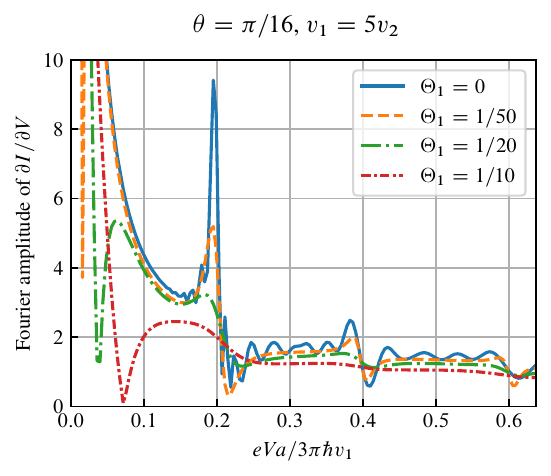}%
	}
	
	\subfloat[\label{subfig:outer_numerical_results:temperature_b}]{%
		\includegraphics[width=.8\columnwidth]{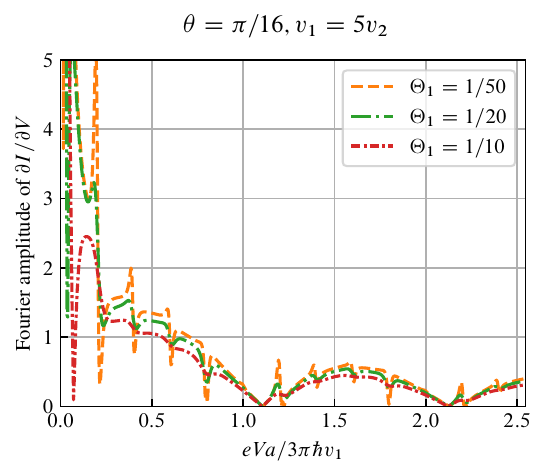}%
	}
	\caption{Outer mode tunneling at $\nu=2/5$ at a fixed island charge. The Fourier amplitude of the interference contribution to the differential conductance is shown at different temperatures, with 
	$\theta=\pi/16$, $v_{1}=5v_{2}$, and $\Theta_{1}=T a/\hbar v_1$. (a) A zoom-in view of the conductance curve for a better visualization of the first few resonances. (b) The conductance curve over a wider range, showing the modulation due to the outer mode. }
	\label{fig:outer_numerical_results:temperature}
\end{figure}

\begin{figure}[!htb]
	\centering
	\includegraphics[width=.8\columnwidth]{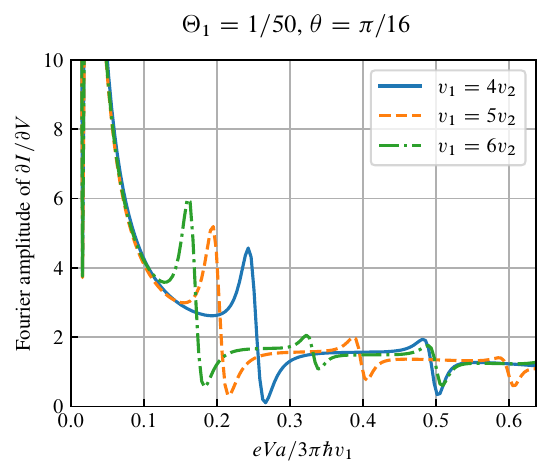}
	\caption{Outer mode tunneling at $\nu=2/5$ at a fixed island charge. Comparison of Fourier amplitudes of outer-mode differential conductance $\partial I/\partial V$ for different velocity ratios $v_1/v_2$, with $\theta=\pi/16$ and $\Theta_{1}=T a/\hbar v_1=1/50$.}
	\label{fig:outer_numerical_results:ratio}
\end{figure}

Figs.~\ref{fig:outer_numerical_results:temperature} and \ref{fig:outer_numerical_results:ratio} show numerical results for the differential conductance of the interference current, with $\theta=\pi/16$ and $v_{1}=5 v_{2}$. 
Immediately, one can notice a striking feature of resonance in these curves when the temperature is low enough.


The first resonance occurs when $ eV a/3\hbar v_{1} \approx \pi /5$, therefore $\hbar \omega_{J} \equiv eV/3 \approx \pi  \hbar v_{1}/5a = \pi \hbar v_{2}/a$. This reminds us of the energy levels of a closed edge mode  $E_n=2\pi \hbar n v/L$, where $n$ is an integer, $v$ is the edge velocity, and $L $ is the circumference of the edge. 
Therefore, the resonance occurs when the Josephson energy $\hbar \omega_{J}$ of the tunneling quasiparticle matches the energy levels of plasmons on the inner mode, which are $n \hbar \pi v_{2}/a$.
As can be seen from Fig.~\ref{fig:outer_numerical_results:temperature}, there is a second resonance when $ eV a/3\hbar v_{1} \approx 2\pi /5$, in agreement with this picture.
One can also check what happens when we change the ratio between $v_{1}$ and $v_{2}$, and the plots are shown in Fig.~\ref{fig:outer_numerical_results:ratio}.
When the ratio $v_{1}/v_{2}$ gets larger, the resonance occurs at a smaller value of $ eV a/3\hbar v_{1}$.
All these findings are consistent with our interpretation.

\begin{figure}[!htb]
	\centering
	\includegraphics[width=.8\columnwidth]{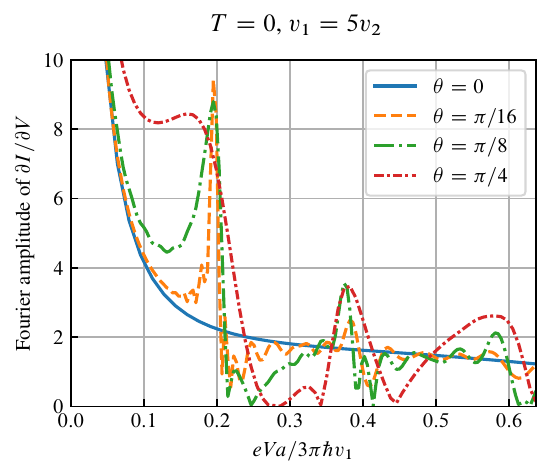}
	\caption{Outer mode tunneling at $\nu=2/5$ at a fixed island charge. The Fourier amplitude of the differential conductance $\partial I/\partial V$ of the interference current is shown for different $\theta$, with $v_{1}=5v_{2}$ and $T=0$.}
	\label{fig:outer_numerical_results:theta}
\end{figure}

There is one issue in our argument:  one can  treat the closed inner $1/15$ mode as an eigenmode only in the weak interaction limit. To find out what happens in the strong interaction limit, we show conductance curves at different $\theta$ in Fig.~\ref{fig:outer_numerical_results:theta}. Clearly, resonance is suppressed at strong interactions.

We give a theory of the resonance feature in the above plots in Appendix~\ref{appendix:resonance}.
We find that at a weak interaction $\theta$ and zero temperature, the width of the resonance scales
as $\theta^2$ and its height scales as $\theta^{-2/3}$. 

\subsubsection{Fluctuating island charge}

\begin{figure}[!htb]
	\centering
	\includegraphics[width=.8\columnwidth]{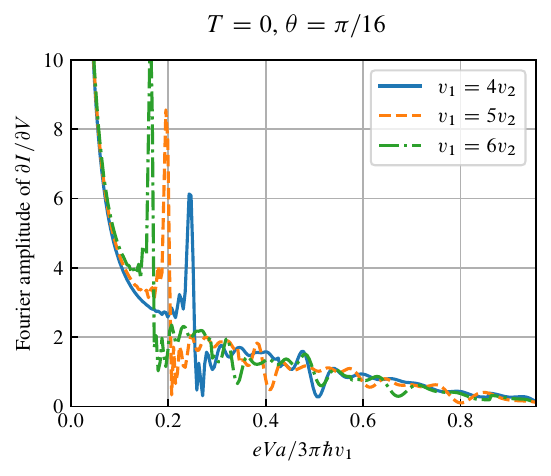}
	\caption{Outer mode tunneling at $\nu=2/5$ for fluctuating island charge. The primary Fourier amplitude of the differential conductance $\partial I/\partial V$ of the interference current is shown for different velocity ratios $v_1/v_2$, with $\theta=\pi/16$ and $T=0$.}
	\label{fig:twofifth_outer_long:ratio}
\end{figure}

\begin{figure}[!htb]
	\centering
	\includegraphics[width=.8\columnwidth]{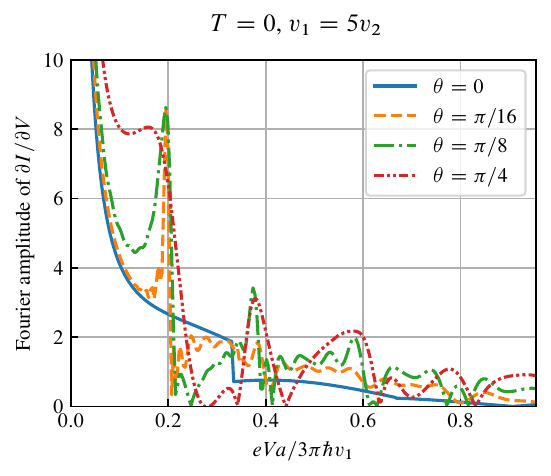}
	\caption{Outer mode tunneling at $\nu=2/5$ for fluctuating island charge. The primary Fourier amplitude of the differential conductance $\partial I/\partial V$ of the interference current is shown for different $\theta$, with $v_{1}=5v_{2}$ and $T=0$.}
	\label{fig:twofifth_outer_long:theta}
\end{figure}

Just like at $\nu=2$, our analysis builds on the case of a fixed island charge and involves averaging the Aharonov-Bohm phase factor over the distribution of the charge of the inner loop.  We use the results of Section \ref{sec:model} and the Fourier expansion from Appendix \ref{appendix:Fourier}. As at $\nu=2$, our results apply only at low temperatures, and they are illustrated in Fig.~\ref{fig:twofifth_outer_long:ratio} and \ref{fig:twofifth_outer_long:theta}. We plot one prominent Fourier harmonic with respect to the magnetic field $\lambda_1$.
As at $\nu=2$, we focus on zero temperature and choose the harmonic of frequency $\alpha_1+ \gamma\beta_1$. This corresponds to the  $m=0$ harmonic from Appendix \ref{appendix:Fourier} and is a natural choice since $\alpha_1\lambda_1$ can be understood as the naive Aharonov-Bohm phase $2\pi\Phi/3\Phi_0$,
where $\Phi$ is the magnetic flux through the nominal area of the device, and $\Phi_0$ is the flux quantum,
and we can interpret $\gamma\lambda_1\beta_1$ as the phase due to the average $\lambda_1$-dependent charge on the inner island. 
The value of $\beta_1$ does not matter for the plots because it only affects the scale of the vertical axis, which is arbitrary here. 
 
In contrast to $\nu=2$, we see two types of sharp features in the differential  conductance:
the resonance from plasmon scattering, addressed in subsection \ref{sec:IV_B_1}, and the voltage-threshold features associated with changes in the fluctuating island charge, similar to the one discussed in subsection \ref{sec:III_C} at $\nu=2$. 
The plasmon resonance manifests itself as a sharp maximum with a nearby minimum on the curve. 
The voltage-threshold feature  is seen as a sharp drop on the curve at $eVa/3\pi\hbar v_1\approx 0.4$ in the blue line in Fig. \ref{fig:twofifth_outer_long:theta}.
Its location can be deduced in the same way as Eq. (\ref{eq:V_nu_2}) using the definitions of $g_{d,u}$, which apply to $\nu=2/5$  in  Appendix  \ref{appendix:additional_phase}. One finds

\begin{equation}
    V=m\frac{\nu_1 w}{2\nu_2 \beta_0 v_{\rm o}},
\end{equation}
where $m$ is an integer, $\nu_1=1/3$, and $\nu_2=1/15$. As at $\nu=2$, resonances allow extracting the parameters of the edge theory from the data.

Another important feature is a rapid decay of the current at a high voltage. As at $\nu=2$, this is a manifestation of the long-time fluctuations of the Aharonov-Bohm phase due to the fluctuations of the charge on the inner loop.
As exceptions, however, 
note a similarity of the red and green curves in Fig.~\ref{fig:twofifth_outer_long:theta} with the red and green curves for a fixed island charge in Fig.~\ref{fig:outer_numerical_results:theta}. It reveals little suppression of interference by the fluctuations of the island charge. The reason is that $\gamma$ in Eq. (\ref{eq:53}) is relatively close to 0 at $\theta=\pi/4$ or $\pi/8$ and $v_1=5v_2$.


\section{Summary}\label{sec:conclusion}

In the main part of this paper, we have used bosonization methods to study nonlinear behavior of 
a quantum Hall Fabry-P{\'e}rot interferometer at finite source-drain voltage, in cases where the quantum Hall state
contains two edge    modes propagating in the same direction.
Important examples are when the bulk filling factor is
$\nu = 2/5$ or $\nu = 2$. We focus on the situations in which the
constrictions defining the interferometer region are either
slightly pinched off, so that the inner mode is weakly
backscattered while the outer mode is completely transmitted, or pinched off to an
extent that the inner mode is completely reflected while
the outer mode is weakly backscattered.  Our analyses
were confined to situations in which the interfering channel is only weakly backscattered, because situations with
stronger backscattering are difficult to treat using the
bosonization approach. 

The most interesting and challenging cases involve tunneling between the outer edge modes in devices where
the inner mode forms a closed loop. In contrast to the
tunneling between inner edge modes, these cases involve
a nontrivial scattering process for bosonic edge plasmons between the opposite edges of the device. After
the scattering problem is solved, however, the residual tunneling
of quasiparticles can be treated perturbatively in the
weak-backscattering limit. If the edge modes are not
too strongly coupled, we find resonance features in the
$I$--$V$ curves at low temperatures, which emerge due to
resonance states of plasmons on the closed inner channel.
The resonances contain information about the velocity of
the inner edge mode. A quantitative analytic theory of
the resonances is possible in the limit of weak intermode
interaction.

Previous investigations \cite{ferraro2017:10.21468/SciPostPhys.3.2.014} have considered the theory for
interferometers at $\nu = 2$, with strongly coupled edge
modes, in the limit where the fast mode velocity is taken
to infinity. Here we have extended the theory to the physically realistic case of finite edge velocities. We have also addressed the effects of asymmetrically-applied voltage bias,
since this is relevant for recent experiments in graphene \cite{Kim}.
Another novel effect addressed in this work is weak tunneling between co-propagating edge channels. While the leakage current through the inner closed loop is negligible in comparison with the total current through the interferometer, fluctuations of the charge on the closed inner channel, caused by the tunneling events, can have
a dramatic effect
on the interference signal at high voltages. Most importantly, the interference contribution to the current rapidly decays as a function of voltage, in agreement with experiments. We have only addressed  these effects at zero temperature, and an extension of the theory to finite temperatures would be useful. 
Indeed, we find that many features of the $I$--$V$ curve are highly sensitive to the temperature in the model, even in the absence of fluctuations of the charge on the inner loop. 

Fluctuations of charge on the inner loop are not
the only mechanism of interference suppression at a high voltage bias. Similar effects can also result from tunneling between the edges and localized states inside the interferometer. Tunneling can also lead to telegraph noise, recently observed in Refs.~\cite{tel1,tel2}, if the tunneling rates are sufficiently slow. This effect can be present even for one-channel edges  if the tunneling particles are anyons or if the bulk-edge interaction is not fully screened. 

We note that in the case of a single channel, at $\nu=1$,  the interference current is predicted to be a periodic function of the bias in a simple model with well-screened bulk-edge interactions \cite{chamon1997:PhysRevB.55.2331}. However, screening gates are typically not close enough to the 2D gas to fully screen the interaction between 
contra-propagating edge channels close to the tunneling contacts. This should make little difference in the limit of weak tunneling at the contacts. If the tunneling probability is on the order of $50\%$, however, the interplay of tunneling and Coulomb interaction can excite plasma modes in the interferometer, which can act like a temperature rise that increases with the voltage and can suppress interference at a high bias. This effect could also be important for interferometers with several edge modes, if backscattering  probabilities at the constrictions are near 50\%. 

We have assumed instantaneous short-range interactions in all our models. This is legitimate if the screening gates are good metals. They can always be treated as such on large time scales, such as the time needed to take a measurement or the time it takes the system to respond to changes in the charge on a closed loop or a localized impurity level.  
However, behavior on the  time scale of  charge flight through the interferometer is more subtle. 
For example \cite{private_communication},   in a recent experiment \cite{nakamura2023:fabry}
at $\nu=2/5$, the screening gates are two-dimensional electron systems with a very large Hall angle, 
$\sigma_{xx}\ll\sigma_{xy}$.
In this case, a more appropriate model might treat the screening layer as one or more additional finite-velocity chiral channels,  interacting with the interferometer modes. We leave an investigation of such a model for future work.

Interesting experimental data have arrived for nonlinear transport in multi-channel quantum Hall interferometers, and new experiments are being performed. We
expect that new data, especially at low temperatures, will
provide an opportunity for fruitful comparison with the
theory. 
We predict resonance features in the weak-coupling regime at $\nu=2/5$, but the temperature was not sufficiently low to observe them in the recent
experiment \cite{nakamura2023:fabry}. 
At the same time, our prediction of rapid suppression of interference at a high voltage bias even for weak tunneling at $\nu=2/5$  agrees with the data
\cite{nakamura2023:fabry}.  

The focus of our theory at $\nu=2$ was on weak tunneling that can be treated perturbatively. 
On the other hand, the experiment of Ref.~\cite{Kim} was carried out in the
50\% transmission regime, and  lower transmission is required for comparison with our work.
Fortunately, the experimental regimes needed for comparison with this paper are within reach.

\acknowledgments
We thank N. Batra, J. Ehrets, P. Kim, M. J. Manfra, J. R. Nakamura, and T. Werkmeister for useful discussions.
This research was supported in part by the National Science Foundation under Grants  DMR-2204635 and DMR-1231319.

\appendix

\section{Series expansion of $P_{\rm o}(t,x)$}\label{appendix:series}

Ref.~\cite{ferraro2017:10.21468/SciPostPhys.3.2.014} provides a technique to represent $P_{\rm o}(t,x)$ as an exponential function of a series, for $\theta = \pi/4$ and $v_{1} \to \infty$. We follow this technique in this appendix, and extend it to an arbitrary $\theta$ and a more general $v_{1}$, as well as to the finite-temperature regime.

\subsection{Zero temperature}

For the purpose of numerical integration, it is better to write $P_{\rm o}(t,a)$ as the product $P_{1}(a) P_{2}(t,a) P_{3}(t)$ of the following three expressions,
\begin{align}
	P_{1}(a) & = \exp \Big( 2 m^2\nu_1 \int_{0}^{\infty} \frac{d\omega}{\omega} 
	S_{11}^{*} 	e^{-\omega \delta}  \Big), \label{eq:P1_zeroT}\\
	P_{2}(t,a) &= \exp\Big[ m^2\nu_1 \int_{0}^{\infty}d{\omega} \frac{1}{\omega} 
	(S_{12} + S_{12}^{*} - 2) e^{-i\omega  t} 
	e^{-\omega \delta}  \Big], \label{eq:P2_zeroT}\\
	P_{3}(t) &= \delta^{-2 m^2\nu_1} \exp\Big[ - 2m^2\nu_1  \int_{0}^{\infty} \frac{d{\omega}}{\omega} 
	(1 -  e^{-i\omega  t } )
	e^{-\omega \delta}  \Big] \nonumber\\
	&= (\delta + it)^{-2 m^2 \nu_1}. \label{eq:P3_zeroT}
\end{align}
We will set $ v_1 = k v_2$, where $k$ is an integer. All our plots correspond to such a choice of the velocities. It is straightforward to generalize to a rational $k$. The technique does not generalize to the case, in which $k$ is an irrational number, but this is irrelevant from a physical point of view.
Now we take the coefficients $S_{11}$ and $S_{12}$ as functions of the dimensionless variable $\omega' = a \omega / v_1$.
Expanding $S_{11}$ and $S_{12}-1$ in series in $\xi=e^{i\omega'}$, we can find
\begin{align}
	S_{11} 	&=  \sum_{n=0}^{\infty} a_{n} (1-\xi) \xi^n, \\
	S_{12}-1 &= \sum_{n=0}^{\infty} b_{n} (1-\xi) \xi^n,
\end{align}
where $a_{n}$ and $b_{n}$ are real numbers.
Next, we will make use of the following formula,
\begin{equation}\label{eq:f_zero}
	f(\eta, A) = \int_{0}^{\infty} \frac{d\omega}{\omega}  (1- e^{-i\omega\eta} ) e^{-\omega A}
	= \ln(1+i \frac{\eta}{A}).
\end{equation} 
Therefore, $P_{1}$ and $P_{2}$ in Eqs.~(\ref{eq:P1_zeroT}) and (\ref{eq:P2_zeroT}) can be expressed in the following form,
\begin{align}
	P_{1}(a) &= \exp\Big\{ 2 m^2 \nu_1 \sum_{n=0}^{\infty} a_{n} f (1, i n + \frac{v_1 \delta}{a}) \Big\} ,\\
	P_{2}(t,a) &= \exp\Big\{ m^2 \nu_1 \sum_{n=0}^{\infty} 
	b_{n}\big[ f(-1, -in +i \frac{v_1 t}{a} + \frac{v_1 \delta}{a}) \nonumber \\
	&~~~~~~~+ f(1, in +i \frac{v_1 t}{a} + \frac{v_1 \delta}{a})\big]
	\Big\}.
\end{align}
For the purpose of numerical computations, we only keep a finite number of terms in these expansions, and the exact number of terms depends on the rate of convergence. 
One can show that the rate of convergence increases as we increase $\theta$.
For the curves shown in Fig.~\ref{fig:outer_numerical_results:theta}, we keep 400 terms for $\theta=\pi/32$, 300 terms for $\theta=\pi/16$, 100 terms for $\theta=\pi/8$, and 80 terms for $\theta=\pi/4$.

\subsection{Finite temperature}
At a finite temperature $T$, $P_{\rm o}(t,a)$ can be written as $P_{1}(a) P_{2}(t,a) P_{3}(t)$ as well.
The finite-temperature expressions of the factors are
\begin{widetext}
	\begin{align}
	P_{1}(a) &= \exp[ 2 m^2 \nu_1  \int_{0}^{\infty} \frac{d\omega}{\omega}
		\left(  \frac{S_{11}^{*}}{1-e^{-\hbar\omega/T}}
		+  \frac{S_{11}}{e^{\hbar\omega/T}-1}
		\right)
		e^{-\omega \delta}  ], \label{eq:P1_finiteT}\\
	P_{2}(t,a) &= \exp[  m^2 \nu_1  \int_{0}^{\infty} \frac{d\omega}{\omega}  
		(S_{12} + S_{12}^{*} -2)  
		\left(\frac{e^{-i\omega t}}{1-e^{-\hbar\omega/T}} +  \frac{e^{i\omega t}}{e^{\hbar\omega/T}-1} \right)
		e^{-\omega\delta}  ], \label{eq:P2_finiteT}\\
	P_{3}(t) &= \delta^{-2 m^2\nu_1} \exp[ - 2 m^2 \nu_1 \int_{0}^{\infty} \frac{d\omega}{\omega}  
		\left(  \frac{1 -  e^{-i\omega t }}{1-e^{-\hbar\omega/T}} + \frac{1 - e^{i\omega t }}{e^{\hbar\omega/T}-1} \right)
		e^{-\omega\delta}  ] = \frac{(\pi T/\hbar)^{2 m \nu_1^2}}{\sin^{ 2 m \nu_1^2}[ (\pi T/\hbar) (\delta + it) ]}.\label{eq:P3_finiteT}
	\end{align}
\end{widetext}

To derive an expansion at a finite temperature, a generalization of Eq.~(\ref{eq:f_zero}) is required. We use the following formula (for a derivation, see Ref.~\cite{vondelft1998:bosonization}),
\begin{align}
	g(\eta, A, \beta) 
	&= \int_{2\pi/L}^{\infty} \frac{d\omega}{\omega}   \left(\frac{e^{-i\omega \eta}}{1-e^{-\beta\omega}} 
		+ \frac{e^{i \omega \eta}}{e^{\beta\omega}-1}\right)  e^{-\omega A} \nonumber\\
	&= - \ln\left\{ \frac{2\beta}{L} \sin\left[ \frac{\pi}{\beta} ( A + i \eta )  \right] \right\},
\end{align}
where the limit $L\to\infty$ is assumed. One can check the agreement with Eq.~(\ref{eq:f_zero}) by computing $g(\eta,A,\beta)-g(0,A,\beta)$ in the limit of $\beta\to\infty$.

We take the coefficients $S_{11}$ and $S_{12}$ as functions of the dimensionless variable $\omega' = a \omega / v_1$ again, but we use a different expansion in $\xi=e^{i\omega'}$,
\begin{align}
	S_{11}  &= \sum_{n=0}^{\infty} c_{n} \xi^n, \label{eq:S11_expansion_finiteT}\\
	S_{12} -1 &= \sum_{n=0}^{\infty} d_{n} \xi^n.
\end{align}
For a side note, one can derive the following relations between $a_{n}, b_{n}$ and $c_{n}, d_{n}$,
\begin{align}
	c_{n} &= a_{n} - a_{n-1},\quad c_{0}=a_{0}, \\
	d_{n} &= b_{n} - b_{n-1},\quad d_{0}=b_{0}.
\end{align}
In Eqs.~(\ref{eq:P1_finiteT}) and (\ref{eq:P2_finiteT}), $P_{1}$ and $P_{2}$ can be represented as
\begin{align}
	P_{1}(a) &= \exp\Big[ 2 m^2 \nu_1 \sum_{n=0}^{\infty} c_{n} g( n ,  \frac{v_{1}\delta}{a}, \frac{\hbar v_{1}}{T a}) \Big],\\
	P_{2}(t,a) &= \exp\Bigg\{ m^2 \nu_1 \sum_{n=0}^{\infty} d_{n}[ g( - n + \frac{v_{1} t}{a} ,  \frac{v_{1}\delta}{a}, \frac{\hbar v_{1}}{T a} ) \nonumber\\
	&~~~~~~~~~~~~~~~~~+ g( n + \frac{v_{1} t}{a} ,  \frac{v_{1}\delta}{a}, \frac{\hbar v_{1}}{T a} ) ] \Bigg\}.
\end{align}

\subsection{Series representation of $P_{\rm o}(t,0)$}
Finally, for the completeness of the discussion, we give an expansion for $P_{\rm o}(t,0)$ as well. At zero temperature, we write $P_{\rm o}(t,0) $ as $P_{3}(t) P_{4}(t)$, where $P_{3}(t)$ is given in Eq.~(\ref{eq:P3_zeroT}) and $P_{4}(t)$ is
\begin{equation}
 \exp\Big\{  m^2\nu_1 \int_{0}^{\infty} \frac{d\omega}{\omega} 
	 ( S_{11}+S_{11}^{*} ) (1-e^{-i\omega t} ) e^{-\omega \delta}  \Big\}.
\end{equation}
We adopt the expansion of $S_{11}$ in the form of Eq.~(\ref{eq:S11_expansion_finiteT}). Therefore $P_{4}$ is represented as
\begin{align}
	\exp\Big\{ &m^2 \nu_1 \sum_{n=0}^{\infty} 
	c_{n}\big[ f(\frac{v_1 t}{a}, -in + \frac{v_1 \delta}{a}) + f(\frac{v_1 t}{a}, in + \frac{v_1 \delta}{a})\big]
	\Big\}.
\end{align}

At a finite temperature, $P_{3}(t)$ is given in Eq.~(\ref{eq:P3_finiteT}), and $P_{4}$ is
\begin{align}
\exp\Bigg\{ & m^2\nu_1 \int_{0}^{\infty} \frac{d\omega}{\omega} 		
	( S_{11}+S_{11}^{*} ) \nonumber\\ 
	&~~~~~\times\left(\frac{1-e^{-i\omega t}}{1-e^{-\hbar\omega/T}} + \frac{1-e^{i\omega t}}{e^{\hbar\omega/T}-1} \right)
	e^{-\omega \delta}  \Bigg\},
\end{align}
which can be represented in the following series form,
\begin{align}
	\exp\Bigg\{ &m^2 \nu_1 \sum_{n=0}^{\infty} c_{n}[ g(0 ,  \frac{v_{1}\delta}{a}-in, \frac{\hbar v_{1}}{T a} ) \nonumber\\
	&-g(\frac{v_{1} t}{a} ,  \frac{v_{1}\delta}{a}-in, \frac{\hbar v_{1}}{T a} ) + g(0 ,  \frac{v_{1}\delta}{a}+in, \frac{\hbar v_{1}}{T a} ) \nonumber\\
	&-g(\frac{v_{1} t}{a},  \frac{v_{1}\delta}{a}+in, \frac{\hbar v_{1}}{T a} )] \Bigg\}.
\end{align}

\section{Effect of asymmetric bias and charge fluctuations on the Aharonov-Bohm phase}\label{appendix:additional_phase}

We need to consider the phase due to the accumulation of charge in the device, %
in addition to the usual Aharonov-Bohm phase. 
This additional phase depends on the bias voltages applied to the lower and upper edges. 
Also, as discussed in Appendix~\ref{appendix:glitch}, it exhibits discrete jumps when the charge on the closed inner channel changes between its allowed quantized values. 
In this Appendix, we explicitly calculate this additional phase for the $\nu=2$ and $\nu=2/5$ interferometers in terms of the bias voltages $V_{d}, V_{u}$, and the total charge of the inner mode $N \nu_{1} e$, where $N$ is an integer. We also determine possible occupation numbers on the closed inner edge as functions of the bias at zero temperature.

The charge density of the right- or left-moving mode at filling factor $\nu$ is $\pm\frac{e \sqrt{\nu} \partial_{x} \phi}{2\pi}$. 
We first write down the Hamiltonian in terms of the original outer and inner modes,
\begin{align}
	H = a \Big[ \frac{\hbar v_{\rm o}}{4\pi} (\partial_{x}\phi_{\rm o}^{u})^2 + \frac{\hbar v_{\rm i}}{4\pi} (\partial_{x}\phi_{\rm i}^{u})^2
	+ \frac{2 \hbar w}{4\pi} (\partial_{x}\phi_{\rm o}^{u})(\partial_{x}\phi_{\rm i}^{u}) \nonumber\\
	+ \frac{\hbar v_{\rm o}}{4\pi} (\partial_{x}\phi_{\rm o}^{d})^2 + \frac{\hbar v_{\rm i}}{4\pi} (\partial_{x}\phi_{\rm i}^{d})^2  + \frac{2\hbar w}{4\pi} (\partial_{x}\phi_{\rm o}^{d})(\partial_{x}\phi_{\rm i}^{d}) \nonumber\\
	- V_{d} \sqrt{\nu_{1}}\frac{e \partial_{x}\phi_{\rm o}^{d}}{2\pi} + V_{u} \sqrt{\nu_{1}} \frac{e \partial_{x}\phi_{\rm o}^{u}}{2\pi}  \Big].
\end{align}
where $\pm \sqrt{\nu_{1}} e\partial_x\phi/2\pi$ stand for average time-independent charge densities, which do not depend on the coordinate on the lower and upper edges.

To find the phase factor, we must know the charge densities in the ground state. Therefore, we minimize the Hamiltonian with respect to the charge densities
~\cite{law2006:PhysRevB.74.045319}. The total charge on the inner channel is $N \nu_{1} e$, where $N$ is an integer. Therefore, we have a constraint
\begin{align}
	\sqrt{\nu_{2}}\frac{e \partial_{x}\phi_{\rm i}^{d}}{2\pi} - \sqrt{\nu_{2}} \frac{e \partial_{x}\phi_{\rm i}^{u}}{2\pi} = \frac{N \nu_{1}e}{a},
\end{align}
where $a$ is the distance between the QPCs.

We first assume a constant $N=0$.
Differentiating the Hamiltonian with respect to the three variables $\partial_{x} \phi_{\rm i}^{u}$, $\partial_{x} \phi_{\rm o}^{u}$, and $\partial_{x} \phi_{\rm o}^{d}$, we find the condition for the extrema
\begin{align}
	\partial_{x} \phi_{\rm i}^{u} &= q_{\rm i}^{u} = -\sqrt{\nu_{1}}\frac{e(V_{d}  -V_{u})  w}{2 \left(v_{\rm i} v_{\rm o}-w^2\right)}, \\
	\partial_{x} \phi_{\rm o}^{u} &= q_{\rm o}^{u} = -\sqrt{\nu_{1}}\frac{ e (2V_{u}v_{\rm i} v_{\rm o}- V_{u}  w^2- V_{d} w^2)}{2 v_{\rm o} \left(v_{\rm i} v_{\rm o}-w^2\right)},\\
	\partial_{x} \phi_{\rm o}^{d} &= q_{\rm o}^{d} = \sqrt{\nu_{1}} \frac{e (2  V_{d} v_{\rm i} v_{\rm o} -  V_{u} w^2 -  V_{d}  w^2)}{2 v_{\rm o} \left(v_{\rm i} v_{\rm o}-w^2\right)}.
\end{align}
To make the ground state expectation values $\langle \partial_{x}\phi_{\rm o/i}^{u/d} \rangle$ vanish, we shift the fields $\phi_{\rm o/i}^{u/d} \to \phi_{\rm o/i}^{u/d} - q_{\rm o/i}^{u/d} x $, and the additional phase factor for $\Gamma_{2}$ is then given by 
\begin{equation}
\label{eq_alpha_0}
	\exp[i\sqrt{\nu_{1}}(q_{\rm o}^{u} - q_{\rm o}^{d}) a]
	= \exp\Big[-\frac{i \nu_{1} e  (V_{d}+V_{u})a}{\hbar v_{\rm o}}\Big]
\end{equation}
This phase depends only on the outer mode velocity $v_{\rm o}$.

Next, we consider the general case when the charge on the inner edge is  $N \nu_{1} e \neq 0$ and induces an additional charge   of $- N (\nu_{1}\sqrt{\nu_{1}}/\nu_{2}) e (w / v_\text{o})$  on the outer edge. Then, the outer-edge phase factor is
\begin{align}
\label{eq_gamma}
	\exp{ \left[-\frac{ie \nu_{1} (V_{d}+V_{u})a}{\hbar v_{\rm o}} + 2\pi i N \frac{\nu_{1}\sqrt{\nu_{1}}}{\sqrt{\nu_{2}}} \frac{w}{v_{\rm o}}\right]}
\end{align}
There is also an additional statistical phase $\exp(-2 \pi i N \nu_{1})$, which matters
for non-integer $\nu_1$ but can be ignored at $\nu=2$.

As an example at filling factor $\nu=2$, when $v_{\rm i} = v_{\rm o}$ and the ratio between the eigen-mode velocities is $v_{1}=5 v_{2}$, from Eq.~(\ref{eq:eigen_velocity}), we can find $v_{\rm o}=v_{\rm i} = 3 v_{2}$ and $w=2 v_{2}$. Therefore, the phase jump due to each additional hole on the inner channel is $+ 4 \pi / 3$, which equals $-2\pi/3 \pmod{2\pi}$.

We now address the dynamics of $N$.
To set up the kinetic equations, we need to count the number of the available states
on the inner loop. To do this, we introduce the chemical potential $eV_{\rm i}$ 
of the inner loop, that is, subtract $a \Big[V_{\rm i} \sqrt{\nu_{2}} \frac{e}{2\pi} \partial_{x} \phi_{\rm i}^{d} - V_{\rm i} \sqrt{\nu_{2}} \frac{e}{2\pi} \partial_{x}\phi_{\rm i}^{u}\Big]$ from the Hamiltonian. Our task is to count available values of the quasiparticle number $N$ on the inner edge for $V_d>V_i>V_u$.
The number of quasiparticles that minimizes the Hamiltonian for a given choice of $V_{\rm i}$ is
\begin{equation}
	N = \left[ \frac{\sqrt{\nu_{2}}}{\nu_{1}} \beta_{0}  \left( \sqrt{\nu_{2}} \frac{2v_o}{w} V_{\rm i} - \sqrt{\nu_{1}} V_{u} - \sqrt{\nu_{1}} V_{d}\right)   \right],
\end{equation}
where $[x]$ is the nearest integer to $x$, and
\begin{equation}
\label{beta0}
	\beta_{0} = \frac{ea}{2\pi\hbar} \frac{w}{v_{\rm i}v_{\rm o} - w^2}.
\end{equation}
Within the above model, the maximal and minimal occupancy of the inner loop can be obtained by substituting
$V_{\rm i}=V_d$ and $V_{\rm i}=V_u$. There are, however, two subtleties, which require a slight modification of our model. The inner loop may couple to various gates. In the simplest model, the electrostatic coupling shifts $V_{\rm i}$ by a constant. A similar shift is needed to accommodate the coupling of the current along the inner loop and the magnetic field \cite{law2006:PhysRevB.74.045319}. 
With this in mind, we set $N=N_{u} \equiv [g_{u}]$ for $V_{\rm i}=V_{u}$ and $N=N_{d}\equiv [g_{d}]$ for $V_{\rm i}=V_{d}$, where the brackets mean the nearest integer,
\begin{align}
	\label{eq:gu}
	g_{u} &= \frac{\sqrt{\nu_{2}}}{\nu_{1}}\beta_{0} \left( \frac{2 v_{\rm o}}{w}  \sqrt{\nu_{2}} V_{u} - \sqrt{\nu_{1}} V_{u} - \sqrt{\nu_{1}}V_{d}  \right) +  \beta_{1} \lambda_{1}, \\
	\label{eq:gd}
	g_{d} &= \frac{\sqrt{\nu_{2}}}{\nu_{1}}\beta_{0} \left( \frac{2 v_{\rm o}}{w} \sqrt{\nu_{2}} V_{d} - \sqrt{\nu_{1}} V_{u} - \sqrt{\nu_{1}} V_{d} \right) +  \beta_{1} \lambda_{1},
\end{align}
and $\lambda_1$ is a control parameter, such as the magnetic flux. $\beta_1$ is a constant that depends on the details of the system. In general, multiple control parameters may be present. We ignore this point to make the notations simpler.
We assume that $V_{d}>V_{u}$ and hence $N_{d}>N_{u}$. Then the range of the possible number of quasiparticles on the inner loop  is between $N_{d}$ and $N_{u}$.

\section{Fourier Expansion with Phase Jumps}
\label{appendix:glitch}

We discuss here the effects of phase jumps arising from an integer constraint on the charge on a closed inner mode on the Fourier expansions of the interference current and the differential conductance of the outer mode. The analysis will be particularly relevant for the case of intermediate coupling between the modes.
We restrict our discussion to the short-time limit, i.e., we assume the island charge is fixed at the value that minimizes the energy for the given value of $\tilde{V}$, with  $V=0$.  Of course, these results apply in the linear transport regime.

We limit our discussion to the short-time limit of the fixed charge of the inner loop at $\nu=2$. It also applies to linear transport.

Assume the effective Aharonov-Bohm phase $\varphi$ can be written as 
\begin{equation}\label{eq:glitch-AB-phase}
	\varphi = C + \alpha_L + P(g) 
\end{equation}
where $C$ is a constant,
\begin{align}
	\alpha_L &= \sum_{j=0}^M \bar{\alpha}_j \lambda_j, \\
	g &= \sum_{j=0}^M \bar{\beta}_j \lambda_j, 
\end{align}
and $P$ is a periodic function of $g$, where $\lambda_j$, for $1 \leq j \leq  M$, are parameters such as the magnetic field and various gate voltages, while $\lambda_0 = \tilde{V}$ .  The coefficients $\bar{\alpha}_j , \bar{\beta}_j$ are constants that depend on the details of the system, and with no loss of generality, we can choose the period of $P$ to be 1.  We are interested here in the linear regime, so we set $V \to 0$.

To describe sharp phase jumps, we choose $P$ to be a sawtooth function,

\begin{equation}
	P(g) =  - \bar{\gamma} ( g - [g] +D),
\end{equation}
where $[g]$ is the closest integer to $g$, and $D$ is a constant. 
Such a choice is justified by the discussion in Section \ref{sec:II_E}.
Without loss of generality, we can set $D=0$ by 
changing the value of $C$.
Then 
\begin{equation}
	P(g) = \sum_n \gamma_n e^{2 \pi i n g} ,
\end{equation}
where $\gamma_n =   - i \bar{ \gamma}  (-1)^n / 2\pi n$ and $\gamma_0=0$.

A similar expansion can be used in a situation where the inner mode is not perfectly reflected at the QPCs. In this case, the  saw-tooth steps will be rounded, and the values of $\gamma_n$ will fall off exponentially at large $n$. 

In the same way, we can write
\begin{equation}
	\label{ft} 
	e^{iP(g)} =\sum_n c_n e^{2  \pi i n g} = \sum_n c_n e^{2  \pi i n (\bar{\beta} _0 \tilde{V} + \sum\bar{\beta}_j \lambda_j)} .
\end{equation}
The coefficients $c_n$ may be expanded in terms of products of $\gamma_n$, if $\bar{ \gamma}$ is small. 

We wish to take the  Fourier transform of $e^{i \varphi}$ with respect to one or more of the parameters $\lambda_j$ for $j \geq 1$.  In the simplest case, we just choose $j=1$, and we hold $\lambda_j$  fixed for all other $j$. (Without loss of generality, we can set these parameters equal to zero by adjusting the constant $C$ and shifting $\lambda_1$.)

Let us define
\begin{equation}
	h(k) =  \int _{-\infty}^{\infty} d \lambda_1 e^{-ik \lambda_1 } e^{i \varphi} 
\end{equation}
We then find that
\begin{equation}
	h(k) = 2 \pi e^{iC} \sum_n  c_n    e^{i \tilde {V} ( \bar{\alpha}_0 + 2 \pi n \bar{\beta}_0)}  \,  \delta (k - \bar{\alpha}_1 - 2 \pi n \bar{\beta}_1 )  .
\end{equation}
We see that  a non-zero value of $\tilde{V}$ leads to  linear shifts in phases  but no change in the magnitudes  of the various Fourier components.

If $\bar{\gamma}=0$, we have 
\begin{equation}
	h (k) =  2 \pi \delta (k - \bar{\alpha}_1) e^{ i \bar{\alpha}_0 \tilde{V}}  e^{iC} 
\end{equation}
More generally, if $\bar{\gamma} = -2 \pi m$, where $m$ is an integer, we have $c_n = 0$ for $n \neq m$, and 
\begin{equation}
	h (k) =  2 \pi \delta (k - \bar{\alpha}_1 - 2 \pi m \bar{\beta}_1 )      e^{ i  \tilde{V} (\bar{\alpha}_0 + 2 \pi m \bar{\beta}_0)  }  e^{iC} .
\end{equation}
If $m$ is not an integer, $h(k)$ will contain multiple Fourier components,  but the largest peak in $h(k)$ will correspond to the integer $n$ that is closest to $m$.

As an example, we consider the case of $\nu=2$, when the parameter $\lambda_1 = B\bar{A}$, where $\bar{A}$ is the nominal area enclosed by the outer edge mode, ignoring the small periodic area oscillations that occur as one varies parameters. Then $\bar{\alpha}_1 =e/\hbar$. We assume that the nominal area enclosed by the inner mode is slightly smaller than $\bar{A}$, and we write it as $ r \bar{A}$, where $r$ is slightly smaller than 1.
Then  $2 \pi \bar{\beta}_1 = r e/\hbar$.  The $n=0$ peak of the Fourier transform then occurs at a frequency $k= e/ \hbar$  while the $n=1$ peak occurs at $k= (1+r) e/\hbar \approx   2e/\hbar$.  In the limit of strong coupling, where $v_1/v_2 \to \infty$, we have $\bar{\gamma} \to  -2 \pi$, and only the $n=1$ peak survives.

Using Eq.~(\ref{basic}) from the main text, we have
\begin{equation}
	\label{IintFourier}
	I_{\text{int}} = \frac { \tilde{I}(V)} {2}  \sum_n [ e^{ i \lambda_1 (\bar{\alpha}_1 + 2 \pi n \bar{\beta}_1)} e^{2 \pi i n \bar{\beta}_0 \tilde{V}} c_n e^{iC} e^{i \bar{\alpha}_0 \tilde {V} }  + \text{c.c.}  ].
\end{equation}
Then
\begin{equation}
	\label{didVn} 
	\frac {d I_{\text{int}}} {dV}  =  \sum_n   [( A_n + A'_n)  e^{ i \lambda_1 (\bar{\alpha}_1 + 2 \pi n \bar{\beta}_1)} e^{2 \pi i n \bar{\beta}_0 \tilde{V}}
	+ \text{c.c.} ],
\end{equation}
where
\begin{align}
	A_n &= \frac{1}{2} \frac {d \tilde {I}} {dV} \,  c_n e^{iC} e^{i \bar{\alpha}_0 \tilde {V} }   ,\\
	A'_n &=  i   \eta (\tilde{I }/2) \,   c_n e^{iC} e^{i \bar{\alpha}_0 \tilde {V} }( 2 \pi n \bar{\beta}_0 + \bar{\alpha}_0) ,
\end{align}
and $\eta = d \tilde{V} / dV$. 
If we calculate  $\int_{-\infty}^{\infty} d \lambda_1  e^{-i k \lambda_1}  (d I_{\text{int}} /dV) $,  the Fourier transform of (\ref{didVn}), we will obtain a set of $\delta$-function peaks at frequencies $ k = k_n^{\pm} = \pm ( \bar{\alpha}_1 + 2 \pi n \bar{\beta}_1) $, with amplitudes given by
$(A_n + A'_n) e^{2 \pi i n \bar{\beta}_0 \tilde {V}}$ or its complex conjugate. 
If $\bar{\alpha}_1$ is not commensurate with $2 \pi \bar{\beta}_1$, these peaks are distinct, and their amplitudes can be measured separately. The magnitudes will be independent of  $\tilde {V}$ and will be given by
\begin{equation}
	\mathcal{A}_n = ( |A_n|^2 + | A'_n|^2 )^{1/2} .
\end{equation}

In our model, at $\nu=2$, if  one is in the strong coupling regime, where $\bar{\gamma}$ is close to $-2 \pi$,  the dominant coefficient $c_n$ will come from $n=1$, and the dominant term  in (\ref{didVn})  will come from $n= 1$. 
If $\bar{\alpha}_1$ and $2 \pi \bar{\beta}_1$ are commensurate,  however, some of the peaks may coincide. Thus, if $r=1$, we have $k_1^+ = k^-_{-3} $, resulting in a single peak with  an amplitude 
$(A_1 + A'_1) e^{2 \pi i  \bar{\beta}_0 \tilde {V}} + (A_{-3}^*  + A'^*_{-3})  e^{6 \pi i  \bar{\beta}_0 \tilde {V}}$.
Then, if $c_{-3}$ is not negligible compared to $c_1$, the  magnitude will depend on $\tilde{V}$ and on the phase offset $e^{iC}$. 
As remarked above, the coefficients other than $c_1$ all vanish in the limit of strong coupling between the modes.

The  discussions above can be readily generalized  to finite temperatures if one replaces $e^{iP(g)}$ in (\ref{ft}) by a thermal average 
$\langle e^{i P(g)} \rangle$ and takes into account the Gaussian phase distribution arising from  thermal fluctuations of the charge on the outer edge. Thermal fluctuations will typically lead to an exponential decrease  of the interference signal at high temperatures, with the weaker Fourier coefficients $c_n$ decaying faster than the dominant one. 
The discussion can also be readily generalized to situations in which one considers two or more parameters $\lambda_j$ and 
takes a multidimensional Fourier transform  with respect to them. 

We finally comment on the slope ${d\varphi}/{d\tilde V}$.
At $\nu=2$,
if $2a$ is taken to be the same for the two modes,  we find  
\begin{equation}
    \bar{\alpha}_0 
    = - \frac {2ea (v_{\text{i}} - w) } { \hbar ( v_{\text{i}} v_{\text{o}} - w^2 )    }
    = 2 \pi \bar{\beta}_0 \frac { v_{\text{i}} -w }
    { v_{\text{o}} -w} ,
\end{equation}
\begin{equation}
    \bar{\gamma} =- 2 \pi w / v_{\text{o}} .
\end{equation}
The slope $d \varphi / d \tilde{V}$ between phase jumps is then given by 
\begin{equation}
    \frac{d \varphi} {d \tilde{V}} =   -2 \frac{ea}{\hbar v_{\text{o}}},
\end{equation} 
as found in Appendix \ref{appendix:additional_phase}. 

By contrast, the slope extracted in an interference experiment from the dominant $n=1$ Fourier peak will be given by
\begin{equation}
    \frac {\Delta \varphi} {\Delta \tilde{V} }=-4 \frac{ea} {\hbar (v_{\text{o}}+w)},
\end{equation} 
as implied by (\ref{IintFourier}). This will be somewhat larger than the differential rate if $w< v_{\text{o}}$. This difference can also be understood from the results of Appendix \ref{appendix:additional_phase}.  When  $\tilde {V}$ is increased by the amount necessary to change  $\varphi$ by $4 \pi$, the occupations of the inner and outer modes will have each increased by 1, so that one phase jump will have occurred. If the negative charge jump on the outer mode has a magnitude less than 1, this will reduce  the  change $\Delta \tilde{V}$ necessary to produce a net increase   of one electron on the outer edge. 

In Section~\ref{sec:II_E}, we averaged the phase factor over the distribution function
$f_{N}$, where $N$ is fluctuating between $N_{d}$ and $N_{u}$. If we assume 
$V_{d}=V_{u}=\tilde{V}$, then $N_{d}=N_{u}$ and $N$ would no longer fluctuate. 
Then, in the notation of Section~\ref{sec:II_E}, the effective phase would reduce  to
\begin{equation}
	\varphi = C + \alpha_{0} \tilde{V}  + \alpha_{1} \lambda_{1} + \gamma N 
    \pmod{2 \pi} ,
\end{equation}
where $N=[g]$ and $g= \kappa \beta_{0} \tilde{V}+\beta_{1} \lambda_1$. For the case 
of $\nu=2$, $\kappa = 2 (v_{\rm o}-w)/w$.
Now $\varphi$ has the same form as in Eq.~(\ref{eq:glitch-AB-phase}), and one can hence make connections between the coefficients $\alpha_{j}, \beta_{j}$ and $\bar{\alpha}_{j}, \bar{\beta}_{j}$. 
By definition, $\gamma = -\bar{\gamma} \pmod{2\pi} $, with  $- \pi < \gamma < \pi$.  We then find the following relations,
\begin{align}
	\bar{\alpha}_{j}  - \bar{\gamma} \bar{\beta}_{j} &= \alpha_{j},\\
	-\bar{\beta}_{0} &= \kappa\beta_{0},\\
	-\bar{\beta}_{1} &= \beta_{1}.
\end{align}%

\section{Effect of fluctuating island charge on the current}\label{appendix:Fourier}
We shall be interested in the Fourier expansion of the following function:
\begin{equation}
	\exp[ i C_{1} F(g+A) + i C_{2} F(g+B)]
	= \sum_{m} a_{m} e^{i 2\pi m g},
\end{equation}
where $A, B, C_{1}, C_{2}$ are constants and $F(g)=g-[g]$. The form of the above function is motivated by Eq. (\ref{average-phase-nu-2}).
The Fourier coefficient $a_{m}$ is given by the formula
\begin{equation}
	a_{m} = \int_{-\frac{1}{2}}^{\frac{1}{2}} d g\, e^{-i2\pi m g} \exp[i C_{1} F(g+A) + i C_{2} F(g+B)].
\end{equation}
We find that $F(g+A)$ can be written as,
\begin{align}
	&F(g+A) = F(g+A-\lfloor A\rfloor) = \nonumber\\
	&\begin{cases}
		g+ A - \lfloor A\rfloor & -\frac{1}{2}<g<\frac{1}{2} - (A-\lfloor A\rfloor)\\
		g+ A - \lfloor A\rfloor -1 & \frac{1}{2} - (A-\lfloor A\rfloor)<g<\frac{1}{2}
	\end{cases}.
\end{align}
In the following calculations, we will use the notation $a\equiv A-\lfloor A\rfloor$ and $b\equiv B-\lfloor B\rfloor$.
We first assume that $a > b$. Then 
\begin{align}
	&C_{1} F(g+A) + C_{2} F(g+B) = \nonumber\\ 
	&\begin{cases}
		C_{1} (g+a) + C_{2} (g+b ) & -\frac{1}{2}<g<\frac{1}{2} - a\\
		C_{1} (g+a-1) + C_{2} (g+b ) & \frac{1}{2} - a<g < \frac{1}{2}-b\\
		C_{1} (g+a-1) + C_{2} (g+b-1 ) & \frac{1}{2} - b<g < \frac{1}{2}\\
	\end{cases}.
\end{align}
Next, $a_{m}$ can be found as
\begin{align}
	a_{m} = {}&\frac{i e^{i C_{1}a+iC_{2}b}}{2\pi m -C_{1}-C_{2}}  e^{-i (2\pi m-C_{1}-C_{2}) /2 }
	\nonumber\\
    &\times [    e^{-i (2\pi m-C_{1}-C_{2}) (-b) } e^{-iC_{1}}(-e^{-i C_{2}}+1) \nonumber\\
	&~~~~+  e^{-i (2\pi m-C_{1}-C_{2}) (-a) } (-e^{-i C_{1}}+1)    ] .
	\label{eq:an_case1}
\end{align}
One can analyze the case of $a<b$ by the substitution $A\leftrightarrow B$ and $C_{1}\leftrightarrow C_{2}$ so that
\begin{align}
	a_{m} = {}&\frac{i e^{i C_{1}a+iC_{2}b}}{2\pi m -C_{1}-C_{2}}  e^{-i (2\pi m-C_{1}-C_{2}) /2 }
	\nonumber\\ 
    &\times [  e^{-i (2\pi m-C_{1}-C_{2}) (-a) } e^{-iC_{2}}(-e^{-i C_{1}}+1) \nonumber\\
	&~~~~+  e^{-i (2\pi m-C_{1}-C_{2}) (-b) } (-e^{-i C_{2}}+1)  ] .
	\label{eq:an_case2}
\end{align}
Due to the definitions of $a$ and $b$, there seems to be an apparent discontinuity in our coefficient $a_{m}$ in Eqs. (\ref{eq:an_case1}) and (\ref{eq:an_case2}) when $A$ or $B$ is an integer. However, this discontinuity is actually absent. Indeed,
in the limit $a \to 1^{-}$, $b$ is less than $a$. We now use (\ref{eq:an_case1}) and find that
\begin{align}
	a_{m} = {}& \frac{i e^{-i (2\pi m-C_{1}-C_{2}) /2 } }{2\pi m -C_{1}-C_{2}} e^{i C_{1}+iC_{2}b}\nonumber\\ 
    &\times   [    e^{-i (2\pi m-C_{1}-C_{2}) (-b) } e^{-iC_{1}}(-e^{-i C_{2}}+1) \nonumber\\
	&~~~~+  e^{i (-C_{1}-C_{2})  } (-e^{-i C_{1}}+1)    ] .
\end{align}
In the limit $a \to 0^{+}$,
\begin{align}
	a_{m} = {}& \frac{i e^{-i (2\pi m-C_{1}-C_{2}) /2 }}{2\pi m -C_{1}-C_{2}} e^{iC_{2}b} \nonumber\\ 
    &\times [   e^{-iC_{2}}(-e^{-i C_{1}}+1) \nonumber\\
	&~~~~+  e^{-i (2\pi m-C_{1}-C_{2}) (-b) } (-e^{-i C_{2}}+1)  ] .
\end{align}
One can see that the two limits are equal, hence no discontinuity occurs.
However, there will generally be discontinuities in the first derivatives with respect to the various parameters.

\begin{widetext}
Eq. (\ref{eq:an_case1}) simplifies in terms of the parameter $\gamma=\frac{2\pi w}{v_o} \pmod{2\pi}$, $C_{1}=-\frac{\gamma}{2} -i\log(\cos{\frac{\gamma}{2}})$, and $C_{2}=-\frac{\gamma}{2} +i\log(\cos{\frac{\gamma}{2}})$ found in Eq.~(\ref{average-phase-nu-2}) as
\begin{align}
	a_{m} &= 
	- \frac{2}{2\pi m + \gamma}  e^{i \pi m (a+b-1)} \cos^{a-b}(\gamma/2) 
	\left\{\sin\left[ \frac{(2\pi m+\gamma)(a-b)-\gamma}{2}\right] + 
    \frac{1}{\cos(\gamma/2)}
    \sin\left[ \frac{(2\pi m+\gamma)(b-a)}{2} \right] \right\} \nonumber\\
	&=  \frac{2 \tan(\gamma/2)}{2\pi m + \gamma}  e^{i \pi m (a+b-1)} \cos^{a-b}(\gamma/2) 
	\cos\left[ \frac{(2\pi m+\gamma)(a-b)-\gamma}{2}\right] .
\end{align}
The result for $a<b$ can be found by exchanging $a \leftrightarrow b$ and $C_{1} \leftrightarrow C_{2}$,
\begin{align}
	a_{m} &= 
	- \frac{2}{2\pi m + \gamma}  e^{i \pi m (a+b-1)} \cos^{a-b}(\gamma/2) 
	\left\{\sin\left[ \frac{(2\pi m+\gamma)(b-a)-\gamma}{2}\right] + \cos(\gamma/2) \sin\left[ \frac{(2\pi m+\gamma)(a-b)}{2} \right] \right\}\nonumber\\
	&= \frac{2\sin(\gamma/2)}{2\pi m + \gamma}  e^{i \pi m (a+b-1)} \cos^{a-b}(\gamma/2)
	\cos\left[\frac{(2\pi m+\gamma)(a-b)}{2}\right].
\end{align}	
\end{widetext}

In the strong coupling regime, at $\nu=2$, where $w/v_{\rm o} \approx 1 $,
and $\gamma=\frac{2\pi w}{v_{\rm o}}\pmod{2\pi}\approx 0$,
we find the largest contribution from $m=0$. (This corresponds to $n=1$ in the notation of the previous Appendix.) Similarly, at $\nu=2/5$, one expects a small $\gamma$ in the strong-coupling regime, where each quasiparticle on the edge of the inner island is screened by a quasihole on the outer edge.

\section{Theory of the resonance}\label{appendix:resonance}
In this section, we provide a quantitative theory of the resonance seen in  Fig.~\ref{fig:outer_numerical_results:temperature}.
We assume weak interaction between the inner and outer edge modes, i.e., $\theta$ is small.
From Eq.~(\ref{eq:P_outer}), we can find the correction $\Delta P$ to $P_{\rm o} (t,a)$ due to this weak interaction. The correction from $\Delta S_{11}$ is merely a constant (independent of $t$) after integration and can be ignored.  The correction due to $\Delta S_{12}$ equals
\begin{equation}
	\Delta P = \frac{1}{3} P_{\rm o}^{0} (t,a)  \int_{0}^{\infty} \frac{d\omega}{\omega} \Delta (S_{12} + S_{12}^{*}) e^{-i\omega t} e^{-\omega\delta} ,
\end{equation}
where $P_{\rm o}^{0} (t,a) = [ \delta + i (t + a/v_{1}) ]^{-1/3 } [ \delta + i (t - a/v_{1}) ]^{-1/3 } $ is the value of $P_{\rm o} (t,a)$ when $\theta =0$.
By looking at Eq.~(\ref{eq:S12}), we can find that
\begin{equation}
	\Delta S_{12} = -\frac
	{2 \theta ^2 \left(e^{\frac{i a \omega  (v_{1}+v_{2})}{v_{1} v_{2}}} \left(\cos \left(\frac{a \omega }{v_{1}}\right)-\cos \left(\frac{a \omega }{v_{2}}\right)\right)\right)+O\left(\theta ^4\right)}
	{\left(-1+e^{\frac{2 i a \omega }{v_{2}}}\right)-2 \theta ^2 \left(e^{\frac{2 i a \omega }{v_{2}}}-e^{\frac{i a \omega  (v_{1}+v_{2})}{v_{1} v_{2}}}\right)+O\left(\theta ^4\right)}.
\end{equation}
To the lowest order, the poles of $\Delta S_{12}$ can be found by calculating the zeros of the denominator, and we can write the poles as $a \omega_{n} / v_{2} = n \pi + \epsilon_{n} + i \Delta_{n} $, where
\begin{equation}
	\label{52}
	\epsilon_{n} + i \Delta_{n} 
	= - i \theta^2 [ 1- (-1)^{n} e^{i n \pi/k} ]
\end{equation} 
with $k = v_{1}/v_{2}$. In this equation, we notice that $\Delta_{n} = -\theta^2 [ 1 -(-1)^{n} \cos(n \pi/k) ] \leq 0 $. 
One needs to go to the next order in $\theta$, if the above imaginary part is zero.
Below we only focus on leading resonances with small $n$, and we assume that the imaginary part of Eq. (\ref{52}) is nonzero.

To find $\Delta P$, we first assume that $t>0$ and consider the contour integral,
\begin{equation}
	\int_{C_{1}+C_{2}+C_{3}} \frac{d z}{z} \Delta (S_{12} + S_{12}^{*}) e^{-iz t} e^{-z\delta} ,
\end{equation}
where $C_{1}: z = s, s \in (0, \infty)$ is the positive real axis, $C_{2}: z = R e^{i\phi}, \phi\in (0, -\pi/2), R\to\infty$ is a quarter circle, and $C_{3}: z = is, s \in (-\infty, 0)$ is the negative imaginary axis. We denote the residue of $\Delta S_{12}$ at the $n$th pole as $M_{n}$, and find the value of the integral as
\begin{equation}
	D_{1}(t)=	2 \pi i \sum_{n} \left(\frac{M_{n} e^{- i \omega_{n} t}}{\omega_{n}}  \right).
\end{equation}
The integral along $C_{2}$ is obviously zero. 
The integral along $C_{3}$ is a featureless function of $t$ since the integrand has no poles close to the imaginary axis. We call that function $h(t)$. 
Notice that $h(t)$ is a real function. Therefore, we can write
\begin{align}
	\int_{0}^{\infty} \frac{d\omega}{\omega} \Delta (S_{12} + S_{12}^{*}) e^{-i\omega t} e^{-\omega\delta} = D_{1}(t) -h(t)
\end{align}
when $t>0$. When $t<0$, we close the contour in the counterclockwise direction and observe that the integral along $C_{1}$ is $D_{2}(t)- h(-t)$, where
\begin{align}
	D_{2}(t) = -2 \pi i \sum_{n} \left(\frac{M_{n} e^{ -i \omega_{n} t}}{\omega_{n}}\right)^{*}   .
\end{align}
For all signs of $t$, we can conclude that the value of the integral is $D(t)-h(|t|)$, where
\begin{equation}
	D(t) =	2 \pi i \sum_{n} \left[\sgn(t) \Re \left(\frac{M_{n} e^{- i \omega_{n} t}}{\omega_{n}}\right) +i  \Im \left(\frac{M_{n} e^{- i \omega_{n} t}}{\omega_{n}} \right)  \right].
\end{equation}

Looking at the formula for the interference current,
\begin{equation}
	\frac{2e^{*}}{\hbar^2} |\Gamma_{1} \Gamma_{2}| \cos\varphi \int_{-\infty}^{0} dt
	\left[ e^{-i\omega_{J}t} P(-t,a) - e^{i\omega_{J}t} P(-t,a) + \text{c.c.}  \right],
\end{equation}
we focus on the correction to the current due to $D(t)$, which can be expressed as
\begin{align}
	&\Delta I = \frac{2e^{*}}{\hbar^2}  |\Gamma_{1} \Gamma_{2}|  \cos\varphi \sum_{n}\int_{-\infty}^{0} dt \nonumber\\
	&\left[ \frac{2 \pi i}{3}  P_{\rm o}^{0} (-t,a)  \frac{M_{n} }{\omega_{n}} \left(e^{-i(\omega_{J}-\omega_{n})t} - e^{i(\omega_{J}+\omega_{n})t} \right) + \text{c.c.}  \right].
\end{align}
We thus expect a resonance when $\omega_{J}$ is close to $v_{2} (n \pi + \epsilon_{n})/a \approx n \pi v_{2}/a$.  The width of the resonance is set by $ v_{2} \Delta_{n} /a \sim \theta^2 $. The height of the conductance maximum can be estimated from the zero bias anomaly for quasiparticle tunneling at $\nu=1/3$. To estimate the height we use the voltage on the order of the resonance width. Then the height is proportional to ${M_{n} \Delta_{n}^{-1/3}}/{\omega_{n}} \sim \theta^{4/3}$.
The conductance peak scales as $M_{n}\Delta_n^{-4/3}\omega_n\sim \theta^{-2/3}$.
All these predictions are in qualitative agreement with the features of the resonance observed in Fig.~\ref{fig:outer_numerical_results:theta}. A quantitative comparison is not possible at large $\theta$.

\section{Table of notations}
\label{appendix:notations}
\begin{table*}[htbp]
    \centering
        \begin{tabular}{ll}
            \midrule
            $a$                                                
            & distance between the two QPCs, measured along an edge  \\
            $\mathcal{A}$                           
            & amplitude of the differential conductance \\
            $\bar{A}$
            & nominal interference area \\
            $B$                                                         
            & magnetic field \\
            $e^*<0$
            & electric charge of the tunneling quasiparticle \\
            $E_{n}$                                                     
            & energy levels of a closed bosonic edge mode\\
            $f_{N}$                           
            & distribution function of the number of inner quasiparticles $N$\\
            $F(g)$                                                     
            & function $g-[g]$, where the bracket $[\dots]$ means the nearest integer \\
            $g_{d}, g_{u}$ 
            & linear functions of controlling parameters $\lambda_{0,1}$ that define $N_{d,u}$ \\
            $G_{d,u}(t,x)$                                              
            & Green's functions of quasiparticle operators on the lower and upper edges\\
            $H$                                                         
            & edge Hamiltonian of the system\\
            $I_{T}$
            & tunneling current \\
            $\tilde{I}$
            & amplitude of the interference current\\
            $L$                                                         
            & edge Lagrangian of the system\\
            $m$                                                         
            & tunneling charge between outer modes in units of $-\nu_1e$ \\
            $n$                                                         
            & tunneling charge between inner modes in units of $-\nu_2e$\\
            $N$
            & net number of positive quasiparticles on a closed inner edge\\
            $N_{d}, N_{u}$
            & maximum and minimum number of inner quasiparticles\\
            $P(t,x)$        
            & Green's function of tunneling operators, defined as $\langle \hat{T}(t,x) \hat{T}^{\dagger}(0,0) \rangle$\\
            $S$                                             
            & scattering matrix from incoming Bose fields to outgoing Bose fields\\
            $\hat{T}(t,x)$                      
            & tunneling operator between lower and upper edges \\
            $T$
            & temperature \\
            $v_{\rm i,o}$, $v_{1, 2}$
            & velocities of various edge modes\\
            $V_{d,u}$ 
            & voltages applied on the lower and upper edges\\
            $V$, $\tilde{V}$ 
            & difference and average of the lower and upper voltages $V_{d,u}$\\
            $w$                                             
            & coupling strength between the outer and inner modes\\
            $\alpha_{0,1}$
            & coefficients of controlling parameters $\lambda_{0,1}$ that directly contribute to the phase $\varphi$\\
            $\beta_{0,1}$
            & coefficients of controlling parameters $\lambda_{0,1}$ in $N$, $N_{d}$ and $N_{u}$ \\
            $\gamma$
            & phase from the interval $(-\pi;+\pi)$ associated with each inner quasiparticle \\
            $\Gamma_{1,2}$                                                 
            & tunneling amplitudes between lower and upper edges at QPC1 and QPC2\\
            $\Gamma_{\rm io}$
            & tunneling amplitude between inner and outer modes on a single edge  \\
            $\delta$                                                 
            & short-time cutoff \\
            $\eta$                                                   
            & asymmetry factor for the voltage bias, defined as $\tilde{V}/V$ \\
            $\theta$                                                 
            & mixing angle of the inner and outer modes\\
            $\Theta_{1,2}$ 
            & dimensionless temperature \\
            $\lambda_{0,1}$                                        
            & controlling parameters in the phase $\varphi$, including the average voltage $\tilde{V}$ and magnetic flux $\bar{A}B$ \\
            $\nu$                                                    
            & filling factor of the bulk region\\
            $\nu_{1}$, $\nu_{2}$
            & filling factors of the outer and inner modes\\
            $\phi_{\rm i, o}$, $\phi_{1,2}$                                         
            & Bose fields of various edge modes\\
            $\varphi$
            & interference phase, i.e., the phase difference between $\Gamma_{1}$ and $\Gamma_{2}$ \\
            $\omega_{J}$                                             
            & Josephson frequency, defined as $e^{*}V/\hbar$ \\
            \bottomrule
        \end{tabular}
\end{table*}

\end{document}